\documentclass[aps, pra, a4paper, showpacs, twocolumn, english, 10pt]{revtex4-1}
\usepackage{bbm, amsthm, bm,textcomp, nicefrac,geometry,ragged2e}
\usepackage[dvipsnames]{xcolor}
\usepackage{float}
\usepackage[bbgreekl]{mathbbol}
\usepackage{graphicx,epstopdf, color, verbatim, enumitem,ulem}
\usepackage{pbox,array}
\usepackage{amsfonts}
\usepackage{amssymb}
\usepackage{mathtools}
\usepackage{multirow}
\usepackage{mathrsfs}
\usepackage{amssymb}
\usepackage{stackrel}
\usepackage[linesnumbered,ruled]{algorithm2e}
\usepackage[thinlines]{easytable}
\usepackage{amsmath}
\usepackage{hyperref}
\hypersetup{
    colorlinks = true,
    linkcolor = Purple,
    citecolor = Purple,
    filecolor = Black,
    urlcolor = Purple,
}
\usepackage{pifont}
\makeatletter
\usepackage[caption=false]{subfig}
\usepackage[T1]{fontenc}
\usepackage[caption=false]{subfig}
\usepackage[english]{babel}
\addto\captionsenglish{}

\newcommand{\be}{\begin{equation}}
\newcommand{\ee}{\end{equation}}
\newcommand{\ba}{\begin{aligned}}
\newcommand{\ea}{\end{aligned}}

\usepackage{lipsum}
\newcommand\id{\mathbbm{1}}
\newcommand{\proj}[1]{\left| #1\right\rangle\!\left\langle#1\right|}
\newcommand{\ketbra}[2]{\left| #1\right\rangle\!\left\langle#2\right|}
\newcommand{\ket}[1]{\left| #1\right\rangle}

\definecolor{brickred}{rgb}{0.8, 0.0, 0.0}

\begin{document}

\title{Nondestructive verification of entangled states via fidelity witnessing}

\author{Ferran Riera-S{\`a}bat}
\thanks{These authors contributed equally to this work.}

\author{Jorge Miguel-Ramiro}
\thanks{These authors contributed equally to this work.}

\author{Wolfgang D\"ur}

\affiliation{Universit\"at Innsbruck, Institut f\"ur Theoretische Physik, Technikerstra{\ss}e 21a, 6020 Innsbruck, Austria}
\date{\today}

\begin{abstract}
Assessing the quality of an ensemble of noisy entangled states is a central task in quantum information processing. Usually, this is done by measuring and hence destroying multiple copies, from which state tomography or fidelity estimation can be employed to characterize states. Here we propose several methods to directly distinguish between two different sets of states, e.g., if their fidelity is above or below a certain threshold value. This turns out to be significantly more efficient and importantly keeps the verified states intact. We make use of auxiliary entanglement or an ensemble of larger size, where we operate on the whole ensemble, but measure only a small fraction where information has been concentrated. For certain state families, we demonstrate that such an approach can even outperform optimal methods that collectively measure directly a fixed fraction of the ensemble.
\end{abstract}

\maketitle

\section{Introduction}
\label{sec:intro}

Quantum entanglement is the key resource for multiple applications in quantum technology, including quantum communication and cryptography \cite{Ekert91, Lo2014}, quantum networks \cite{Wehner_2018, Pirker_2019, Azuma_2021}, distributed computation \cite{CiracDistributed, Cacciapuoti2020, Hayashi15} and distributed sensing \cite{Sekatski2020, Kessler2014, Eldredge2018}. However, in any realistic scenario channels and devices are imperfect, and resulting states will be noisy. Assessing whether the quality of produced or maintained entangled states is sufficient for the desired application is hence a central task for all these applications.

The typical approach to determine the quality of an ensemble of mixed states is to perform local measurements on some individual copies, from which features of the states can be assessed. Multiple strategies for certifying entangled states have been proposed \cite{Eisert2020, Bdescu2019, Thinh_2020, Yu_review2022, Kliesch2021}. The approaches differ in the amount of information learned from the states, which is generally related to the amount of resources spent in the process. The existing strategies range from a complete characterization of the states via state tomography \cite{Cramer2010, Haah_2017}, to the learning of some specific properties such as the fidelity as in fidelity estimation \cite{Flammia2011}. In this work we consider a still less information-demanding problem, consisting in determining whether the fidelity of an entangled state is above or below a certain threshold value, a natural and realistic extension of the state verification problem \cite{Wang2019, Li19, Zhu1, Zhu2, Zhu19, Yu_2019, Hayashi_2009}. We denote this decision problem as fidelity witnessing, which can be relevant in multiple scenarios, whenever it suffices to know if the quality of the states of an ensemble is large enough to perform the desired task. For instance, in a communication scenario, the fidelity directly indicates the error rate of transmitted quantum information via teleportation, but also if the states can be used to expand a secure secret key. The main advantage of such an approach lies in the reduced required resources as compared to schemes that acquire more information.

Here, in contrast to many previous approaches \cite{Yu_2019, Yu_review2022}, we assume that we have access to the whole ensemble, and not just individual copies that are processed locally and individually, and also consider entanglement-assisted protocols. While this poses additional experimental challenges, we find that these approaches have significant advantages: First, they are more efficient and can lead to up to exponential enhancements as compared to previous schemes, even outperforming optimal, non-local methods that measure all states from an ensemble of fixed size. Second, certified states are not destroyed; the remaining states are directly certified, without the necessity to assume a tensor product structure of the initial ensemble. The key element is an information transfer from multiple copies of the states in the ensemble to a certain subset, or to some auxiliary entangled states, where only these few copies are subsequently measured and hence destroyed. In this way, a much larger fraction of an ensemble can be left intact, while still deciding if the remaining states are suitable for the desired application. We also take care of properly accounting for additional entanglement resources, by relating auxiliary entanglement to the required number of noisy copies.

We introduce three different protocols, which are compared with the typically considered case of sequential measurements of single copies of identical, noisy entangled states $\rho^{\otimes n}$. We propose strategies that rely on collective operations which allow us to transfer information about the noise of several copies into a few auxiliary state(s), in the spirit of \cite{Riera1, miguel2022improving}. By partially or completely measuring the auxiliary system(s), one can access the information of the accumulated noise with increased efficiency, in such a way that the copies that are certified are not consumed in the process. We make use of this powerful tool and introduce several approaches, each of which works better in specific situations. To this aim we consider different state families, including noisy ensembles that result from maximally entangled states affected by decay noise, phase-flip noise or depolarizing noise, mimicking relevant situations such as distributing entanglement through noisy channels or storing entangled states for some time in an imperfect quantum memory. The main findings can be summarized as follows:
\begin{itemize}
    \item We introduce three different protocols for the decision problem of distinguishing between two state families of noisy entangled states with respect to their fidelity.
    \item We demonstrate that auxiliary entanglement can be used to increase the performance of this task.
    \item We show that collective but local operations performed on the whole ensemble can be used to transfer and concentrate information into a few copies, thereby significantly reducing the number of states that need to be measured and hence destroyed, while maintaining more, and fully certified states.
    \item We find that for decay noise, our protocols perform exponentially better than optimal, global and collective strategies that only operate on ensembles of a fixed size where all states are measured. 
\end{itemize}

The paper is organized as follows. We review some basic concepts and operations in Sec.~\ref{sec:background}, while the problem setting is formally defined in Sec.~\ref{sec:setting}. In Sec.~\ref{sec:strategies} we introduce the different strategies we propose to solve the fidelity witnessing problem, where we also analyze and compare their efficiency and performance. Finally, we present some concluding remarks in Sec.~\ref{sec:conclusions}. 

\section{Background}
\label{sec:background}
We discuss here the basic notions and operations we make use of throughout this work.

\subsection{Maximally entangled states}
\label{sec:bell:states}

\textit{Bell states.} Bell states are maximally entangled quantum states shared by two qubits. The set of Bell states form a basis of $\mathbb{C}^2_A \otimes \mathbb{C}^2_B$ given by the elements

\begin{equation*}
    \ket{\Psi_{ij}}_{AB} \equiv \id \otimes \sigma^j_x \, \sigma^i_z \, \left( \frac{ \ket{00}_{AB} + \ket{11}_{AB} }{\sqrt{2}} \right),
\end{equation*}
where $i$, $j \in \{0, 1\}$ and $\sigma_k$ is the $k^{\text{th}}$ Pauli operator. These states are a fundamental resource for multiple applications such as, e.g., super-dense coding \cite{Liu2002}, quantum teleportation \cite{Bennett1993, Bouwmeester1997}, QKD \cite{Ekert91, Lo2014} or distributed quantum computation \cite{CiracDistributed, Cacciapuoti2020}.

\textit{Higher dimensional maximally entangled states.} Qudits are natural extensions of qubit systems for $d$-dimensional systems. A bipartite system of qudits is associated with the Hilbert space $\mathcal{H}_{AB} = \mathbb{C}_A^d\otimes\mathbb{C}_B^d$, where we can define an orthonormal basis of maximally entangled states of the form \cite{Bennett1993}
\begin{equation}
    \label{eq:generalized:bell}
    \ket{\Phi^d_{mn}}_{AB} \equiv \frac{1}{\sqrt{d}} \sum_{k = 0}^{d-1} e^{i 2 \pi km/d} \ket{ k }_A \ket{ k \ominus n }_B,
\end{equation}
where $m$, $n \in \mathbb{Z}_d$ are called the phase and amplitude index of the state respectively, and where $k\ominus n\equiv (k-n)\text{mod}\,d$ and $d$ is the dimension of the qudit systems.

\subsection{Fidelity of quantum states}
\label{sec:states:fidelity}

The fidelity of two quantum states $\rho$, $\sigma$ is a measure of how close the two states are, with respect to the probability of identifying one as the other with an optimal measurement. Formally, it is defined as 

\begin{equation}
    \label{eq:fid}
    F = \left[ \text{tr} \sqrt{ \sqrt{\rho} \sigma \sqrt{\rho} } \, \right]^2.
\end{equation}
We use fidelity as a natural figure of merit to measure the amount of noise affecting a quantum state. Since we mainly deal with maximally entangled states, the fidelity in our case refers to the distance between some state $\rho$ and the maximally entangled state $\ket{\Psi_{00}}$, such that Eq.~\eqref{eq:fid} reduces to $F = \left\langle \Psi_{00} \right| \rho \left|\Psi_{00} \right\rangle$.

\subsection{Families of states}
\label{sec:depolarization}

We introduce here the different families of states we make use of throughout this work as probe states to be verified or witnessed. All of them correspond to dominant noise processes relevant in many physical scenarios.

\subsubsection{Bell diagonal states}

Any arbitrary bipartite mixed state $\rho$ can be always depolarized into a state diagonal in the Bell basis with local operations. This is achieved by implementing a quantum map with Kraus operators $\mathcal{D}_{\text{BD}} = \left\{ \frac{1}{2} \, \sigma_i \otimes \sigma_i \right\}_{i=0}^3$, i.e.,
\begin{equation*}
\begin{aligned}
    \mathcal{D}_{\text{BD}} \! : \; & \rho = \sum_{i_1, j_1, i_2, j_2 = 0}^1 \alpha_{i_1 j_1 i_2 j_2} \ketbra{\Psi_{i_1 j_1}}{\Psi_{i_2 j_2}} \\
    \mapsto \; & \rho_{\text{BD}} = \sum_{i,j = 0}^1 \alpha_{ijij} \proj{\Psi_{ij}}.
\end{aligned}
\end{equation*}
Note that the fidelity of the state remains unchanged.

\subsubsection{Werner states}

Further depolarization is possible by making all but one of the diagonal elements equal and transforming the state into a Werner state \cite{Werner89}, i.e.,
\begin{equation}
    \label{eq:werner}
    \rho_{\text{w}} = q \proj{\Psi_{00} } + \frac{1-q}{4} \id_4,
\end{equation}
while keeping the fidelity $F=(3q+1)/4$ unchanged. This is accomplished by suitable twirling techniques (see, e.g., \cite{horodecki}), i.e.,
\begin{equation*}
    \mathcal{D}_\text{w} \! : \, \rho \, \mapsto \int \left( U \otimes U \right) \, \rho \, \left( U \otimes U \right)^\dagger \, \mathrm{d} U = \rho_{\text{w}},
\end{equation*}
where $\mathrm{d} U$ is the Haar measure. Depolarization can also be achieved using only local operations drawn from a discrete set \cite{Bennett1996}. Werner state can be conceived as the worst case in terms of the noise of the states, where with some probability no information about the state is left. 

\subsubsection{Dephasing type states}

A less general family of states is given by rank-2 Bell diagonal states. These states are local unitary (LU) equivalent to a Bell state, where one of the parties is affected by a bit-flip noise, i.e.,
\begin{equation}
\begin{aligned}
   \label{eq:depashing}
   & \id \otimes \mathcal{N}_F \! : \, \proj{\Psi_{00}} 
   \\ & \mapsto \rho_{\text{d}} = F \proj{\Psi_{00}} + (1-F) \proj{\Psi_{10}},
\end{aligned}
\end{equation}
where $\mathcal{N}_p = \{\sqrt{p} \, \id, \, \sqrt{1-p} \, Z \}$. Notice that states resulting from local dephasing noise are formally equivalent, where both qubits may be affected by Pauli $\sigma_z$ noise. This kind of noise is relevant, e.g., when there are fluctuating fields or phase references.

\subsubsection{Amplitude damping type states}

A different class of states corresponds to the amplitude damping type. These states are the result of sending each party of a perfect Bell state $\ket{\Psi_{11}}$ through an amplitude damping channel with Kraus operators $\mathcal{A}_p = \{ \proj{0} + \sqrt{p} \, \proj{1}, \, \sqrt{1 - p} \, \ketbra{0}{1} \}$ \cite{nielsen_chuang_2010}, i.e.,
\begin{equation*}
\begin{aligned}
    & \mathcal{A}_F \otimes \mathcal{A}_F \! : \, \proj{\Psi_{11}} \\
    & \mapsto \rho = F \proj{\Psi_{11}} + (1-F) \proj{00},
\end{aligned}
\end{equation*}
where the state is LU equivalent to
\begin{equation}
    \label{eq:dampingstates}
    \rho_a = F \proj{\Psi_{00}} + (1-F) \proj{01}.
\end{equation}
This kind of state is relevant in scenarios where one describes decay processes of, e.g., atoms in a quantum memory.

\subsection{The counter gate}
\label{sec:counter:gate}

In several of the strategies introduced in this work, we make use of a quantum gate introduced in \cite{Riera1, Riera2} which allows us to transfer information from an ensemble of entangled qubit states into a higher-dimensional entangled state by means of local operations. The so-called counter gate \cite{Riera1, Riera2} is a bilateral qubit-qudit controlled operation that takes a two-dimensional entangled state as control and a $d$-dimensional entangled state as a target. 
Given a target system consisting of a maximally entangled state with phase index zero, see [Eq.~\eqref{eq:generalized:bell}], its action is given by
\begin{equation}
    \label{eq:acction:CX}
    b\text{CX}^{AB}_{1\to 2} \ket{mn}_{1} \ket{\Phi_{0j}^d}_2 = \left|mn\right\rangle_{1} \ket{\Phi^d_{0, j\ominus m \oplus n}}_2,
\end{equation}
where $\ket{mn}$ are the computational basis states, $b \text{CX}^{(d)}_{1 \to 2} = \text{CX}_{1 \rightarrow 2}^{A_1 A_2} \otimes \text{CX}_{1 \rightarrow 2}^{B_1 B_2}$, and
\begin{equation*}
    \text{CX}_{1\rightarrow 2} = \left| 0 \right\rangle\left\langle 0 \right| \otimes \id_d + \left| 0 \right\rangle\left\langle 1 \right| \otimes X_d
\end{equation*}
is the hybrid \textit{controlled-X} gate \cite{Daboul2003}, where the action of $X_d$ in the computational basis is given by $X_d \ket{k} = \ket{ k \ominus 1}$. We denote as type-1, type-2, and type-3 error states, the states corresponding to $\ket{01}$, $\ket{10}$, and $\ket{\Psi_{10}}$ respectively. The action of the counter gate taking a type-1 (-2) error state acting as control leads to an amplitude index value of the auxiliary state increased(decreased) by one, whereas it is left invariant for the type-3 error state.

We denote the operation consisting in applying the counter gate, Eq.~\eqref{eq:acction:CX}, from each of the states of an ensemble of $n$ copies into a $d$-dimensional auxiliary state, as error number gate (ENG), i.e.,
\begin{equation}
    \label{eq:ENG}
    \text{ENG} = \prod_{k=1}^n b\text{CX}_{k\to \text{aux}}.
\end{equation}
The name is motivated by the action of the gate on ensembles with only $|01\rangle$ error states, where the number of error states in the ensemble can be determined in this way. 

\subsection{Entanglement cost and relation to resources}
\label{sec:entanglement:cost}

In this work, we consider entanglement-assisted protocols that make use of (small) amounts of extra entanglement. In order to make a comparison with protocols that only measure states directly, we need to relate these entangled states with the states of the noisy ensemble. First, we point out that the amount of entanglement that is contained in a noiseless Bell state constitutes the basic unit of bipartite entanglement, usually denoted as \textit{ebit} of entanglement. The number of ebits contained in a maximally entangled state of dimension $d$ is given by $E(\ket{\Phi_{00}^d}) = \log_2 d$. Therefore, in an ensemble of $n$ maximally entangled states the number of ebits is given by $E\big( \ket{\Phi_{00}^d}^{\otimes n} \big) = n \log_2 d$.

The evaluation of the amount of entanglement, in terms of the number of ebits, contained in mixed states is however not clear. The distillable entanglement $E_D$ \cite{Horodecki09}, i.e., the fraction of maximally entanglement states that can be distilled from many noisy copies by means of local operations and classical communication is a suitable entanglement measure in this context. It tells us that from $m$ noisy copies of the state $\rho$, one can generate $m E_D$ ebits of entanglement. This provides the desired relation between noisy copies and ebits. However, $E_D$ is hard to compute in general, and only some upper bounds \cite{Vedral1997, Bennett_hashing} and lower bounds \cite{Pirandola_2017} are known. Here we make use of reachable lower bounds provided by entanglement purification protocols \cite{Bennett1996, Deutsch1996, Bennett_hashing, Duur_2007}. We compute the yield $Y$ of a combination of the recurrence protocol \cite{Deutsch1996} and hashing \cite{Bennett_hashing}, i.e., the fraction maximally entangled states over initial noisy states, which is a lower bound for $E_D$ \cite{Hostens2006}. While the recurrence protocol allows one to increase the fidelity of remaining pairs, hashing is only applicable for sufficiently high fidelities, but has a non-zero yield. Both protocols operate on states diagonal in the Bell basis, where any state can be transformed to such form as discussed above. The yield of the hashing protocols for Bell diagonal states $\rho_{BD}$ is given by $[1-S(\rho_{\text{BD}})]$, and one can obtain a smooth curve as a function of the initial fidelity $F$ by mixing strategies, i.e., consider a mixture of states resulting from a certain number of rounds of the recurrence protocol, and apply Hashing to the resulting one.

So whenever auxiliary entanglement of $m$ ebits is required, we assume that it is generated from the noisy ensemble by consuming $\lceil m/Y \rceil$ states from the ensemble, where $Y$ is the yield of the entanglement purification protocol. Notice that we do not optimize over entanglement purification protocols, so we provide a conservative bound, thereby underestimating the operational advantage of entanglement-assisted protocols.

%{\wduer that is diagonal in the Bell basis}, in the asymptotic limit $n \to \infty$, one can distill $n [1-S(\rho_{\text{BD}})]$ maximally entangled states \cite{Bennett_hashing, Riera1, Riera2}, where $\rho_{\text{BD}} = \mathcal{D}^{(d)}_1 (\rho)$ and $S(\rho) = \text{tr}(\rho \log_d \rho)$; and therefore we can assume that a general ensemble of $n$ bipartite qudit mixed states, {\wduer that are diagonal in the basis $\{ \ket{\Phi_{ij}^d} \}$} contains
%\begin{equation}
%    E \left( \rho^{\otimes n} \right) = n [ 1 - S(\rho_{\text{BD}}) ] \log_2 d
%\end{equation}
%ebits of entanglement.

%---------------------------------------------------------------------------------------

\section{Problem setting}
\label{sec:setting}

Quantum state certification and verification have been studied and analyzed in different directions, ranging from learning more information about a state, as in state tomography, to less information as in fidelity estimation. The problems we study here are closely related to quantum state verification of entangled states \cite{Pallister18, Bdescu2019, Thinh_2020, Yu_review2022, miguel2022improving}, where the task is to verify whether some ensemble of entangled states consisting of $n$ identical copies is indeed maximally entangled. In terms of the fidelity of the states, state verification solves the problem of deciding whether the fidelity of the states is $F = \left\langle \Psi_{00}\right|\rho\left|\Psi_{00}\right\rangle = 1$ or not (i.e., $F \leq 1 - \epsilon$).

\begin{figure}
    \centering
    \includegraphics[width=\columnwidth]{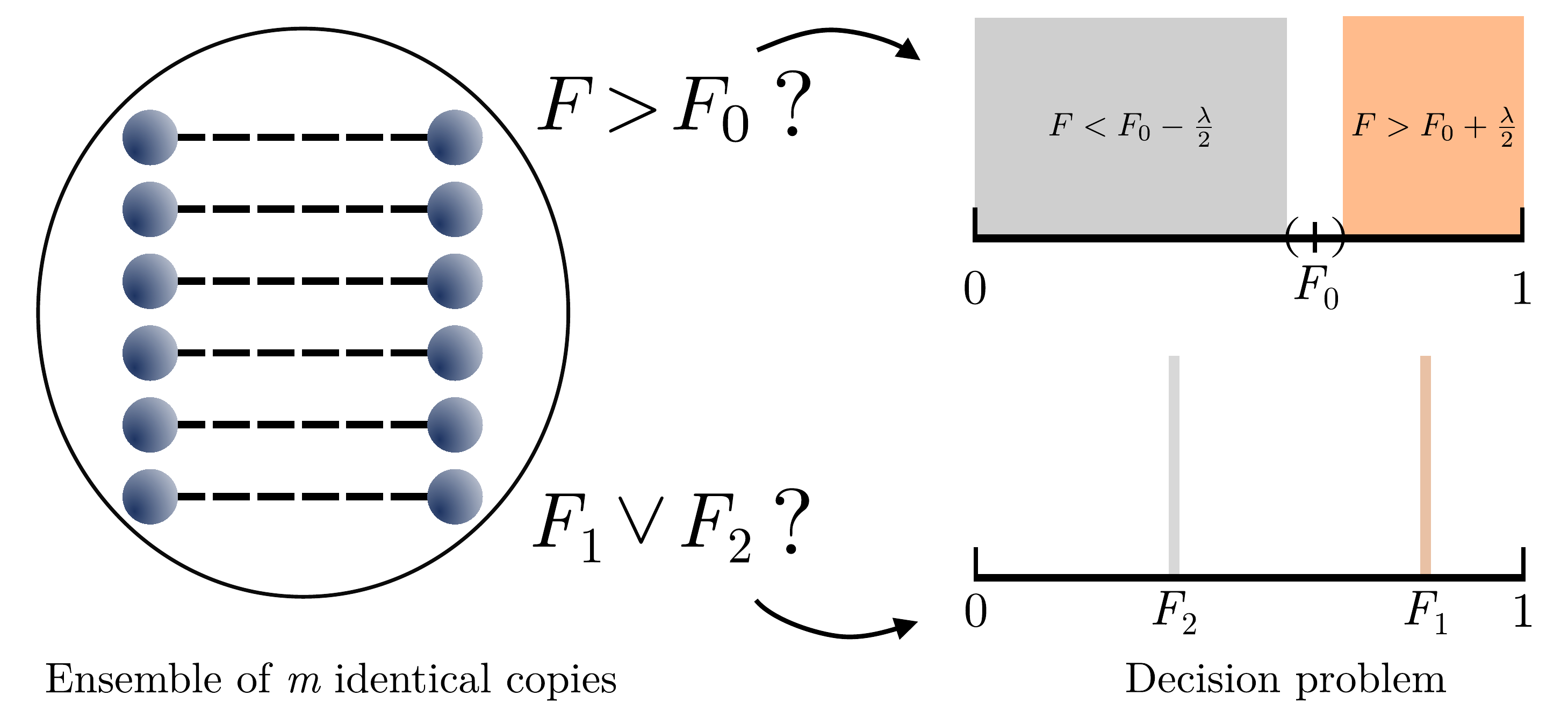}
    \caption{Fidelity witnessing (upper right) and fidelity discrimination (lower right) problems. Given an ensemble of identical copies, $\rho^{\otimes n}$, with some unknown local fidelity $F$ (i.e., the fidelity of each copy, corresponding to their reduced density operator), the task of fidelity witnessing is to determine whether the fidelity is below or above certain threshold $F_0$ up to some additive error $\lambda$. Similarly, if there exists the promise that the ensemble states fidelity is either $F_1$ or $F_2$, the task of fidelity discrimination reduces to deciding which of the two cases is present.}
    \label{fig:witnessingeneral}
\end{figure}

The problems we analyze enclose more realistic situations \cite{Bdescu2019, Yu_2019, Thinh_2020, Yu_review2022}, where a certain level of noise is in general unavoidable, and therefore entangled states cannot be strictly verified with previous methods. The task reduces to a decision problem of determining whether the fidelity of an entangled state is above or below some threshold value, or corresponds to one out of two possible values. We define and analyze these two closely related problems, denoted as fidelity witnessing and fidelity discrimination (see Fig.~\ref{fig:witnessingeneral}).

Notice that it is not required to learn the actual value of $F$ to solve the underlying decision problem which essentially requires only learning one bit of information. This learning of a minimum amount of information can be enough and useful in many contexts and applications, for which the parties sharing the entanglement just require a minimum fidelity of the states to operate. In contrast to other methods, e.g., fidelity estimation or state tomography, fidelity witnessing and fidelity discrimination have the benefit of a significant reduction in the amount of consumed resources. We construct strategies that avoid measuring and hence destroying copies of the state in order to obtain information about it. As we show later, directly measuring copies of the state generally provides additional information that is not required to solve the witnessing problem, making such a strategy wasteful in terms of the required resources.

We consider an ensemble of identical copies of some noisy state, i.e., $\rho^{\otimes n}$, where we can collectively operate on the copies in a local way. All the strategies we propose and analyze rely on the only use of local operations, such that each state is not accessible or operated in a global way. We remark that our approaches do not only restrict to ensembles of the form $\rho^{\otimes n}$, but may also be extended to work with, e.g., non-IID ensemble states.

\textit{Fidelity witnessing.} The witnessing problem consists in determining, with only the assistance of local operations and classical communication, if the fidelity of a certain noisy entangled state $\rho$ is above or below a specific threshold value $F_0$, i.e., $F > F_{0} + \frac{\lambda}{2}$ or $F < F_{0} - \frac{\lambda}{2}$, up to some additive error $\lambda$ (see Fig.~\ref{fig:witnessingeneral}). This problem has been partially analyzed in the context of extending quantum state verification to more realistic settings \cite{Wang2019, Li19, Zhu1, Zhu2, Zhu19, Yu_2019}.

Observe that there exist four different regimes or scenarios that need to be analyzed with any approach. On the one hand, if a protocol output determines that the fidelity is below the threshold, there is some probability of succeeding, i.e., the fidelity is actually below, or failing. Analogously, the other two regimes consist of the protocol determining that the fidelity is above and being right or failing. Each strategy we analyze can exhibit better performance, sometimes tunable, on some of these regimes at the expense of the others.

\textit{Fidelity discrimination.} We also analyze a slightly modified problem, denoted as fidelity discrimination. Given an ensemble of identical entangled states shared by parties $A$ and $B$, with the promise that the fidelity of the states is either $F_1$ or $F_2$, with $F_1 > F_2$, the task is to discern which of the two fidelities correspond to the ensemble copies by consuming the minimum number of states (see Fig.~\ref{fig:witnessingeneral}). Since this task also entails a decision problem with simpler promises, all the protocols we introduce in the following are directly applicable with, in general, enhanced efficiency. In particular, the blocking strategy (protocol P3, see below) allows us to overcome optimal bounds in solving the fidelity discrimination problem.

%-------------------------------------------------------------------------------------------------------------------------------------------------
\section{Fidelity witnessing strategies}
\label{sec:strategies}

We propose different approaches to solve the fidelity witnessing and discrimination problems for ensembles of identical bipartite entangled states. While the most direct approach, a direct extension of strategies applied in fidelity estimation and verification \cite{Flammia2011, Pallister18}, relies on a copy-by-copy measurement of a certain part of the ensemble, the remaining approaches make use of different tools in order to transfer information about the noise of the ensemble into auxiliary registers without destroying the original copies. The auxiliary registers are subsequently manipulated to extract the required information and solve the decision problems. 

We remark that the protocols we introduce are suited for different state families. While protocol P0 (individual measurements) and P3 (blocking strategies) are suitable for all families, the error-counting protocol P1 and the coarse-graining protocol P2, are essentially only applicable to states of the from $\rho_a$ resulting from decay or amplitude damping. 

\subsection{Protocol P0: Individual measurements}
\label{sec:protocol0}

Reference \cite{Yu_2019} discusses how to perform fidelity witnessing by extending the strategy for verification of Bell states.

\begin{figure}
    \centering
    \subfloat[\centering]{\includegraphics[width=0.5\columnwidth]{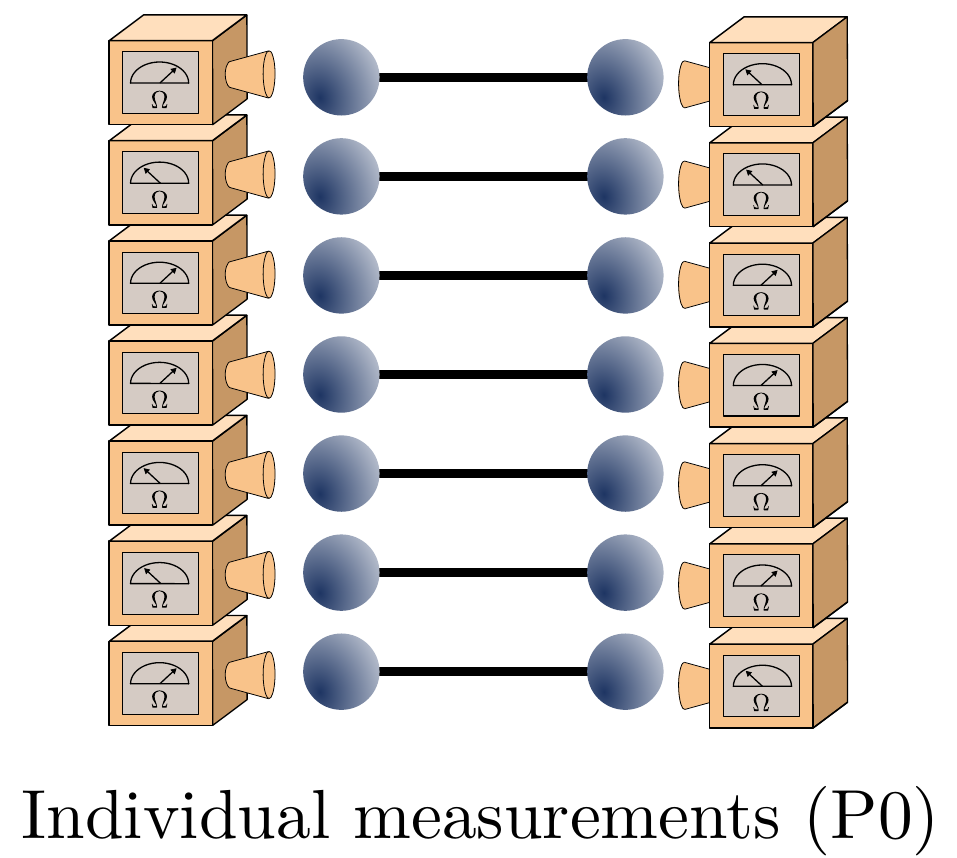} \label{fig:P0}}
    \subfloat[\centering]{\includegraphics[width=0.5\columnwidth]{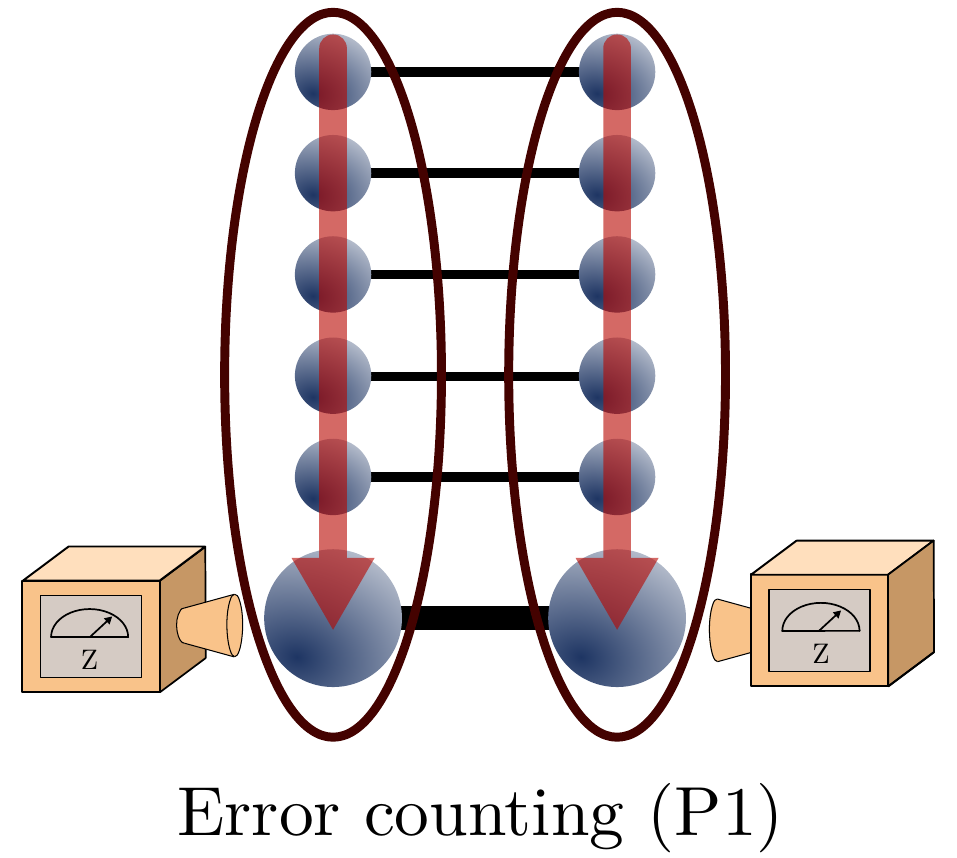} \label{fig:P1}} 
    \caption{\label{fig:schemesP0P1} (a) Schematic representation of the individual measurements protocol P0, which consist in individually measuring each of the copies. (b) An illustrative description of the underlying idea behind the strategies we introduce, P1, P2 and P3, where some information of the ensemble is transferred into an auxiliary state that is then manipulated to learn the required information without destroying the verified ensemble copies. }
\end{figure}

In state verification \cite{Wang2019, Li19, Zhu1, Zhu2, Zhu19, Yu_2019, Hayashi_2009}, an arbitrary noisy ensemble is given and one aims to distinguish between either the states being perfect, i.e., $\rho = \proj{\Psi_{00}}$, or being noisy up to some fidelity, i.e., $\left\langle \Psi_{00} \right| \rho \left| \Psi_{00} \right\rangle \leq 1-\epsilon \equiv F$. The optimal solution in this case consists in performing random copy-by-copy and local two-outcome measurements [see Fig.~\ref{fig:P0}], given by $\{ \Omega_i, \id - \Omega_i \}$, such that
\begin{equation*}
    \Omega_i \ket{\Psi_{00}} = \ket{\Psi_{00}},
\end{equation*}
where the label ``pass'' (``fail'') is assigned to the outcome corresponding to $\Omega_i$ ($\id-\Omega_i$). The measurement can then be described by
\begin{equation*}
    \Omega = \sum_i p_i \, \Omega_i
\end{equation*}
where $\{p_i\}$ is a probability distribution. The probability that one state passes the measurement test is given by
\begin{equation*}
    p = 1 - \epsilon \, \nu(\Omega),
\end{equation*}
where $\nu(\Omega)$ is the spectral gap between the largest and the second largest eigenvalues of $\Omega$. Therefore, for an unknown state $\rho$ the optimal strategy is such that maximizes $\nu(\Omega)$. For Bell states $\ket{\Psi_{00}}$ it is shown \cite{Pallister18} that $\max_{\Omega} \nu(\Omega) = 2/3$ and hence $p_{\max} = (1+2F)/3$.

After measuring $n$ copies, the probability of finding $j$ states that ``fail'' the test and $n-j$ states ``pass'' it is given by
\begin{equation}
    \label{eq:binomial1}
    \text{Pr}(j|F) = \binom{n}{j} \, p^{n-j} (1-p)^j,
\end{equation}
where $F$ is the fidelity of the states. Thus, the value of $j$ is related to the actual fidelity of the ensemble. We exploit this property by suitably acquiring and processing information about the value of $j$ in order to tackle the fidelity witnessing and discrimination problems. We detail later in this section the strategy based on single copy measurements for these two problems. Before, we particularize the previous properties for two types of states that we make use of throughout the manuscript.

\textit{Amplitude-damping noise ensemble.} Consider the states of the ensemble to be amplitude-damping type [Eq.~\eqref{eq:dampingstates}] with some unknown fidelity $F$. By measuring the state on the $Z_A \otimes Z_B$ basis, we consider the state ``pass'' the measurement test if these two outcomes coincide. This is given by
\begin{equation*}
    p = F < p_{\max}.
\end{equation*}
Since we can distinguish with total certainty between the two eigenvectors (with no-null eigenvalue) of $\rho$, and this measurement is equivalent to a measurement with $\nu = 1$, what has been shown to be optimal for the verification problem \cite{Pallister18}. 

\textit{Werner-type states.} Consider now the case where the states of the ensemble are given by Werner states [see Eq.~\eqref{eq:werner}]. Note that this corresponds to depolarizing noise acting on the states. In particular, any state can be brought to a Werner-type form by depolarization means \cite{Werner89}, and hence this case represents a completely general situation.

The optimal measurement in this case also corresponds to locally measuring the states in the computational basis, i.e., with respect to the observable $Z_A$ and $Z_B$, considering that the state ``pass'' the measurement test if the two outcomes coincide. However, the probability of a state with fidelity $F$ passing the measurement is now given by $p = (1+2F)/3$, which is equivalent to performing the optimal measurement for an unknown state. Note that the process is analogous to first depolarizing the state to its Werner form and then performing the described measurement.

\textit{Fidelity witnessing.} Given an ensemble of entangled states with fidelity $F$ and some threshold fidelity $F_0$, the expected value of $j$ (see above) is given by
\begin{equation*}
    \langle \, j \, \rangle = n (1 - F) \nu (\Omega).
\end{equation*}
This allows us to extend this strategy to perform fidelity witnessing. After performing the measurement $\{ \Omega, \id - \Omega \}$ on $n$ copies, we can conclude that the fidelity lies on one or another regime, i.e.,
\begin{equation}
\label{eq:criteri0}
\begin{cases}
    F > F_0 \quad & \text{if} \quad j < n (1 - F_0) \nu (\Omega) , \\
    F < F_0 \quad & \text{if} \quad j > n (1 - F_0) \nu (\Omega).
\end{cases}
\end{equation}
Since the ensemble is in a tensor product structure, the success probability $P_s$ of solving the witnessing problem based on Eq.~\eqref{eq:criteri0}, can be determined by the Chernoff–Hoeffding theorem \cite{Chernoff_1952}
\begin{equation*}
    P_s \geq 1 - e^{ D ( j || n p ) },
\end{equation*}
where
\begin{equation*}
    D( S || Q ) = \sum_{x} S(x) \log \left( \frac{S(x)}{Q(x)} \right)
\end{equation*}
is the Kullback–Leibler divergence (see \cite{Yu_2019}).

The criterion given by Eq.~\eqref{eq:criteri0} can be generalized to tune the success probability as a function of $F$. To show this, we use a different approach. As a starting point, $F$ is given by some prior probability $\varrho(F)$. However, as $j$ and $F$ are not independent random variables, once the value of $j$ is obtained, the probability of $F$ is then given by $\varrho(F|j)$. From the new probability distribution, we make a statement about $F$. In particular, if we define as $\Sigma$ the set of values of $j$ for which we conclude that the fidelity is above the threshold $F_0$, $\Sigma$ contains the values of $j$ for which the probability of the conditioned probability of $F>F_0$ is larger than a parameter $\delta$, which we can freely choose. Formally, we define $\Sigma$ as
\begin{equation}
    \label{eq:e:area}
    \Sigma = \bigg\{ j \, \bigg| \, \delta < \text{Pr} (F>F_0 | \, j) =
    \int_{F_0}^1 \varrho( F| j) \, \mathrm{d} F \bigg\},
\end{equation}
where we use Bayes' theorem to compute the conditioned density probability, i.e.,
\begin{equation*}
     \varrho(F|j) = \frac{\text{Pr}(j|F) \, \varrho(F)}{\int_0^1 \text{Pr}(j|F') \, \varrho(F')  \, \mathrm{d} F'}.
\end{equation*}
Considering a flat probability distribution for $F$, i.e., $\varrho(F) = \text{const.}$, we obtain
\begin{equation*}
    \varrho(F|j) = \frac{\text{Pr}(j|F)}{1+n} .
\end{equation*}

For a given fidelity $F$, the success probability, i.e., the probability of deciding that the fidelity is above (or below) the threshold and being right, is given by
\begin{equation*}
    \label{eq:cases}
    P_s(F) = 
    \begin{cases}
    \sum_{j \not{\in} \Sigma} \text{Pr}(j|F) & \text{for } F < F_0 \\
    \sum_{j \in \Sigma} \text{Pr}(j|F) & \text{for } F > F_0.
    \end{cases}
\end{equation*}
\begin{equation*}
\end{equation*}
The value of $\delta$ can be tuned in order to optimize the performance in certain regimes. Observe that, by appropriately choosing the value of $\delta$ in Eq.~\eqref{eq:e:area}, one can increase or decrease the success probability $P_s$ of the protocol within some of the regimes discussed in Sec.~\ref{sec:setting}, i.e., either enhance the success probability to correctly certify states with $F > F_0$, or correctly reject states with $F<F_0$. We analyze the performance of this protocol in detail in the following sections, comparing it with the other strategies we propose.

\textit{Fidelity discrimination.} The information learning process is analogous for fidelity discrimination, where one learns the number $j$ of states that ``fail'' the measurement test $\{ \Omega, \id - \Omega \}$. Once a particular value of $j$ is obtained, one simply concludes that the fidelity of the ensemble is $F_1$ in case $\text{Pr}(j|F_1) > \text{Pr}(j|F_2)$. 

Similar as before, we separate all the possible values of $j$ in two sets $\Sigma_1$ and $\Sigma_2$, such that
\begin{equation}
\label{eq:discrimination1}
\begin{aligned}
    \Sigma_1 = \left\{ j \, | \, \text{Pr}(j|F_1) > \text{Pr}(j|F_2) \right\}, \\
    \Sigma_2 = \left\{ j \, | \, \text{Pr}(j|F_2) > \text{Pr}(j|F_1) \right\}.
\end{aligned}
\end{equation}
In case fidelity $F_i$ is then given with probability $\eta_i$, the success probability reads as
\begin{equation*}
    P_s = \eta_1 \sum_{j \in \Sigma_1} \text{Pr} (j | F_1) + \eta_2 \sum_{j \in \Sigma_2 } \text{Pr} (j | F_2).
\end{equation*}
If the two fidelities are given with the prior probability $\eta_1 = \eta_2$, the success probability $P_s$ is then bounded by
\begin{equation}
    \label{eq:tracedis}
    P_s \leq \frac{1}{2} \left[ 1 + T \! \left( \rho_1^{\otimes n}, \rho_2^{\otimes n} \right) \right],
\end{equation}
where $\rho_i = \rho(F_i)$, and $T(\rho,\sigma)$ is the trace distance between states $\rho$ and $\sigma$ \cite{nielsen_chuang_2010}. In particular, for amplitude damping noise (see above), the success probability when performing fidelity discrimination reaches the optimal values, i.e., it saturates the trace distance between the two ensembles [Eq.~\eqref{eq:tracedis}].

%--------------------------------------------------------------------------------------------------------------------------------------------------------

\subsection{Protocol P1: Error counting}
\label{sec:protocol1}

The first protocol we propose is an extension of the approach for quantum state verification we introduced in \cite{miguel2022improving}. In the following, we will consider states of the form $\rho_a$ resulting from amplitude damping. In general, given an ensemble of $n$ identical copies of states $\rho_a$, the protocol consists in applying the ENG gate, Eq.~\eqref{eq:ENG}, between $n$ copies of the ensemble and an auxiliary $d$-level system of the form $\ket{\Phi^d_{00}}$ [see Eq.~\eqref{eq:generalized:bell}]. The ENG transfers information about the number of type-1 and type-2 errors in the ensemble into the auxiliary system [see Fig.~\ref{fig:P1}] by changing its amplitude index. By subsequently measuring this auxiliary state we can learn its amplitude index $j$, revealing information about the number of errors, and therefore the fidelity, of the ensemble. Crucially, the remaining ensemble that is witnessed is not destroyed in the process. Notice, though, that the state of the ensemble changes, and is no longer of tensor product structure, i.e., the copies are not independent. However, the fidelity of individual reduced states is known, and above the threshold value $F_0$ in case of successful certification. We detail in the following, the steps for the different cases depending on the kind of noise affecting the ensemble copies.

\textit{Fidelity witnessing.} Consider first that the noise affecting the states of the ensemble is amplitude damping type Eq.~\eqref{eq:dampingstates}. An ensemble of $n$ copies of states of the form Eq.~\eqref{eq:dampingstates} is given. Note that one can also interpret an ensemble of $n$ states of this form as an unknown distribution of pure states where each state corresponds either to $\ket{\Psi_{00}}$ or $\ket{{01}}$. The joint state of the ensemble can then be written as
\begin{equation*}
    \rho^{\otimes n} = \sum_{j=0}^n \binom{n}{j} F^{n-j} (1-F)^j \, \Gamma_j,
\end{equation*}
where $\Gamma_j$ is a density operator corresponding to all permutations of $\{ \ket{\Psi_{00}}_{AB}^{ \otimes (n-j) } \ket{01}_{AB}^{\otimes j} \}$. 

Information about the noise of the ensemble can be transferred and accumulated in a single $d$-level state. For that end, a maximally entangled state of dimension $d=n+1$ is prepared, $\ket{\Phi_{00}^d}$ (see Eq.~\eqref{eq:generalized:bell}). The ENG gate, Eq.~\eqref{eq:ENG}, is applied between the states of the ensemble and the auxiliary $d$-dimensional state, such that

\begin{equation*}
\begin{aligned}
    \text{ENG:} & \; \rho^{\otimes n} \otimes \proj{\Phi_{00}^d} \\
    \mapsto \, & \sum_{j=0}^n \binom{n}{j} F^{n-j}(1-F)^j \, \Gamma_j \otimes \proj{\Phi_{0j}^d}.
\end{aligned}
\end{equation*}
Observe that, given the nature of the noise and the effect of the ENG gate, the amplitude index of the auxiliary state $j$ encodes very specific information about the number of errors contained in the ensemble, that directly relates to the fidelity of the copies.

\begin{figure}
    \subfloat[\centering]{\includegraphics[width=0.5\columnwidth]{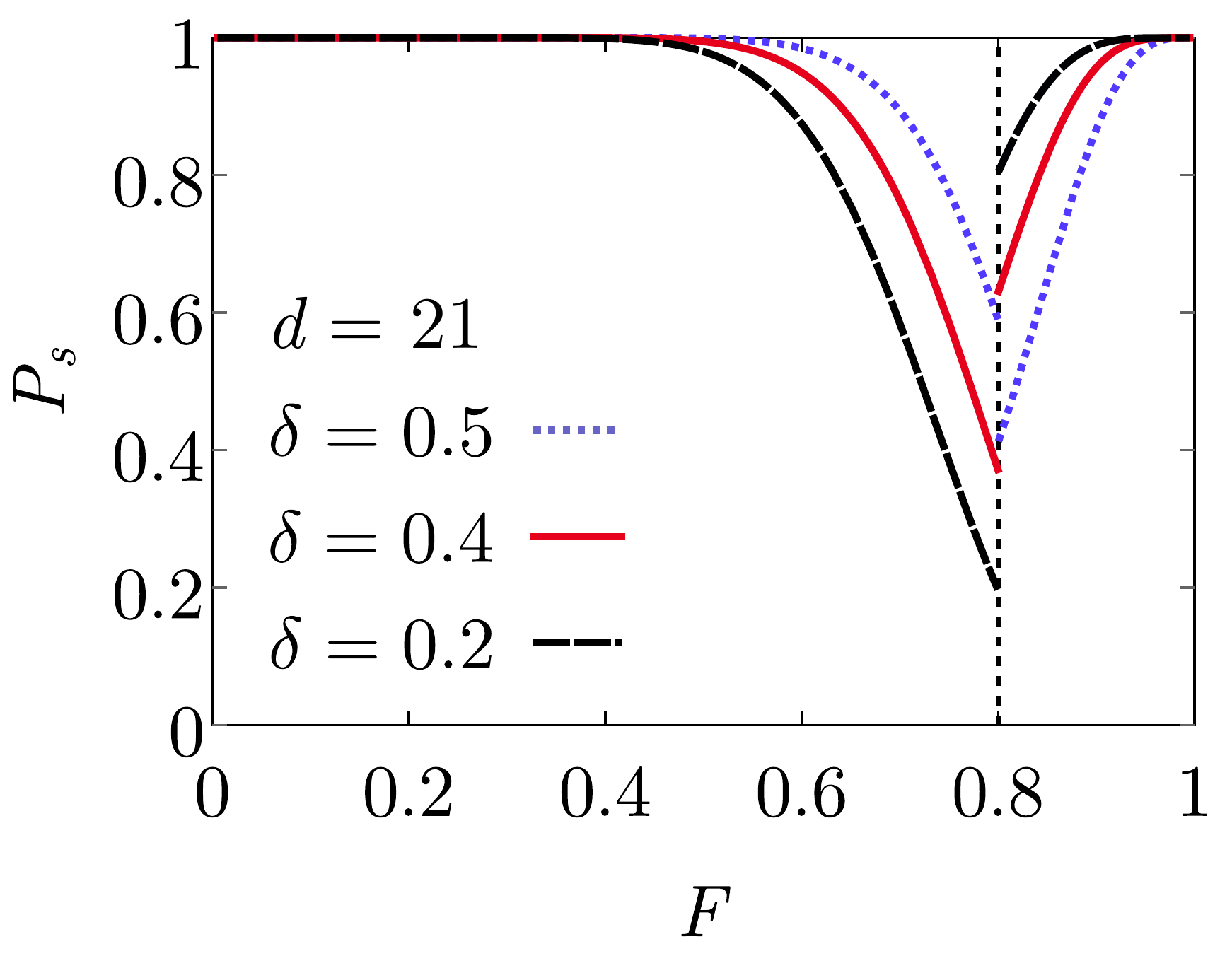} \label{fig3a}}
    \subfloat[\centering]{\includegraphics[width=0.5\columnwidth]{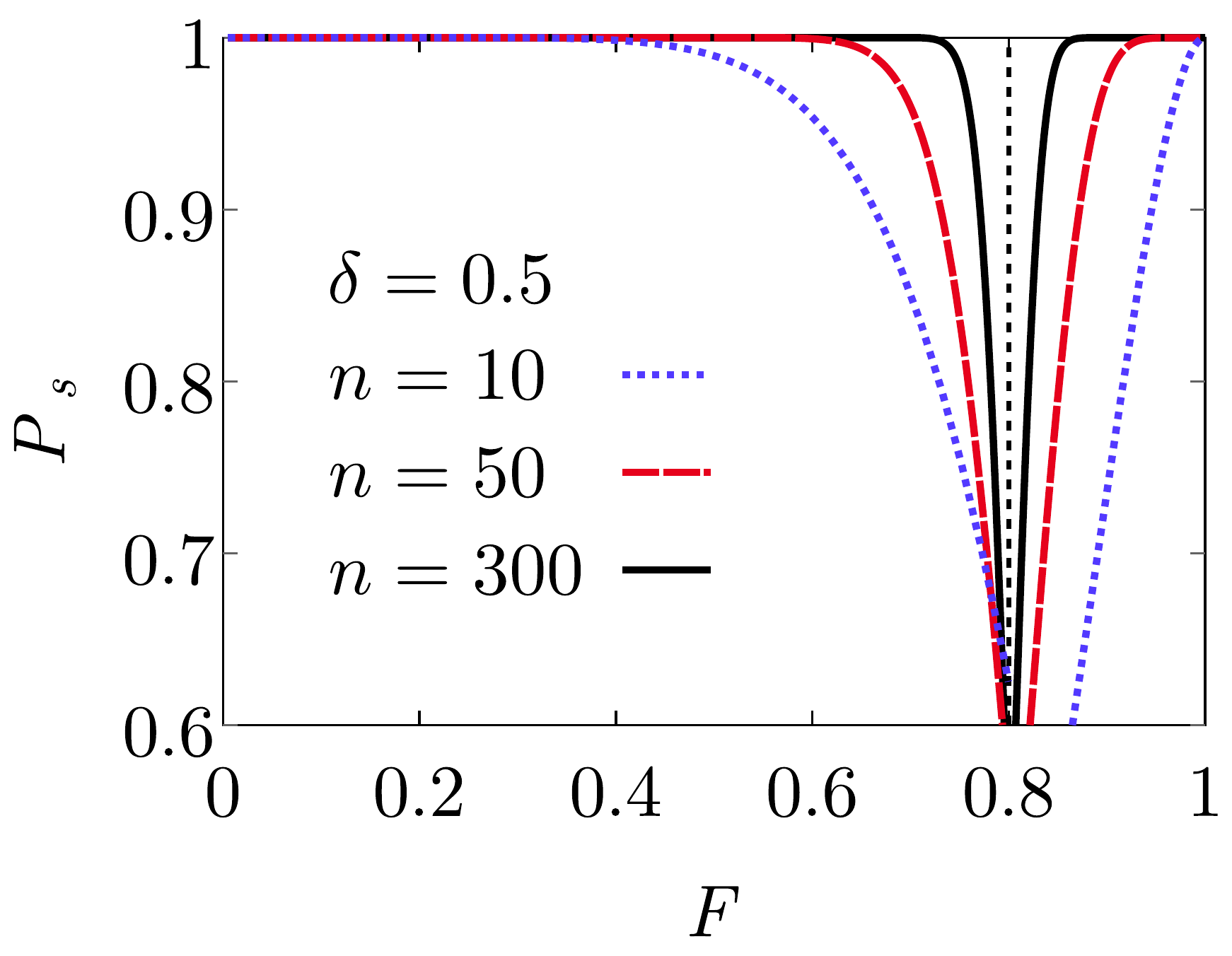} \label{fig3b}}
    \caption{\label{fig3} Performance of the individual measurements (P0) and the error counting (P1) strategies. (a) Shows the success probability of both protocols for an ensemble of $n = 20$ identical copies and different values of the heuristic $\delta$ value, as a function of the actual fidelity of the copies. Depending on the choice of $\delta$, a better performance is found in one or another regime. (b) Shows the success probability of both protocols with a fixed value of $\delta = 0.5$ and different values of ensemble size $n$, as a function of the actual fidelity of the copies. A larger ensemble provides better performance.}
\end{figure}

By simply measuring the auxiliary state locally by parties $A$ and $B$ in the generalized $Z$ basis, and subtracting their outcomes, one learns the value of the amplitude index $j$, from which one can infer the number of errors in the ensemble. Each $j$ value can be found with probability 
\begin{equation*}
    \text{Pr}(j|F) = \binom{n}{j} F^{n-j} (1-F)^j.
\end{equation*}
Note that this corresponds to the same distribution obtained with protocol P0, see Eq.~\eqref{eq:binomial1}, with $p=F$. Due to this close, but probabilistic, dependence, one can solve the decision problem by determining whether the measured value of $j$ is above or below a certain value $j_0 = n(1-F_0)$, which also depends on the threshold problem fidelity $F_0$. Note that the same statistical analysis performed in protocol P0 can be applied here, and hence, the two protocols exhibit the same success probability, see Fig.~\ref{fig3}. However, while in protocol P0 the whole ensemble is consumed, here we just destroy an auxiliary maximally entangled state of dimension $d = n+1$, providing an exponential improvement in the amount of consumed resources. Importantly, the states that are witnessed or verified are not destroyed in the process, in opposition to protocol P0 and previous approaches. Notice, however, that the remaining ensemble is no longer of tensor product structure, but corresponds to $\Gamma_j$ when obtaining the outcome $j$. The fidelity of reduced states, i.e., tracing out all but one copy, is given by $F=(n-j)/n$, which can be larger than the initial fidelity $F$. 

\begin{table}[h]
    {\LinesNumberedHidden
    \label{table:P1}
    \begin{algorithm}[H]
        \SetKwInOut{Input}{Input}
        \SetKwInOut{Output}{Output}
        \SetAlgorithmName{Protocol P$\!\!$}{}
        \justifying \textit{Input}: Ensemble of $n$ identical noisy Bell states and auxiliary $n+1$-level maximally entangled state.
        \begin{enumerate}
            \item Apply the ENG between the ensemble and the auxiliary system. 
            \item Obtain the amplitude index $j$ by measuring locally in $Z$-basis. 
            \item If $j \in \Sigma$ output $F>F_0$, otherwise output $F<F_0$.
        \end{enumerate}
    \justifying \textit{Output}: The fidelity of the initial ensemble was above or below $F_0$ with some success probability $P_s$.
    \caption{Error counting}
    \end{algorithm}}
\end{table}

The protocol can equivalently be applied to Werner states of the form Eq.~\eqref{eq:werner}. The steps of the protocol are identical as before. However, the ensemble can now contain the three different kinds of errors, and hence, the value of $j$ after the application of the ENG does not encode the number of errors in the ensemble. Instead, it provides information about the difference between type-1 and type-2 error states. In this case, the probability of obtaining $j$ for a given $F$, corresponds to the sum of all configurations with a difference of $j$ errors, i.e., 

\begin{equation*}
    \text{Pr}(j|F) = \sum_{\substack{i, k, l = 0 \\ i + k + l = n \\ k \ominus l = j } }^n \frac{n!}{i! \, k! \, l!} \, A^i \left( 1-A \right)^{k + l},
\end{equation*}
where $A \equiv (1+2F)/3$.

Note that, due to the different effect of the counter gate in this case (Sec.~\ref{sec:counter:gate}), we can now obtain $2n +1$ different values of $j$, i.e., $j \in \{ -n, \dots, n \}$. In order to differentiate between all these possible values of $j$, we require an auxiliary state of dimension $d=2n+1$.
We remark, though, that we did not succeed in using this to obtain an efficient protocol for witnessing Werner states, i.e., a protocol that outperforms P0.

\textit{Fidelity discrimination.} The error counting protocol can be directly applied for solving the fidelity discrimination problem. This is accomplished by simply processing the information learnt about the value of $j$ as in the protocol P0 case [see Eqs. \eqref{eq:discrimination1} and \eqref{eq:tracedis}], finding comparable efficiency.

\subsection{Protocol P2: Coarse graining}

In the strategies introduced above, i.e., the single copy measurement and the error counting protocols, the information obtained about the ensemble is unnecessarily excessive. Instead of learning information about whether the number of errors in the ensemble is above or below some value, we, indirectly, also obtain information about the concrete number of errors in the ensemble. This implies that the protocols likely spend more resources than the ones required to just solve the witnessing and discrimination problems.

We propose here a modified protocol that tries to minimize the amount of obtained information, therefore reducing the amount of resources spent during the process. The strategy relies on the application of coarse-grained techniques that allow us to locally access partial information about the amplitude index $j$ of the previous protocol, without destroying the auxiliary state nor the ensemble copies.

\textit{Fidelity witnessing.} We restrict to the amplitude damping noise case for simplicity, i.e., states of the form $\rho_a$. The first steps of the protocol are identical to the error-counting approach. A $d$-dimensional auxiliary state $A_{1}B_{1}$ is prepared in the state $\ket{\Phi_{00}^d}$ with $d=n+1$, and the ENG operation is applied from $n$ states of the ensemble to the auxiliary, such that the information of the number of errors of the ensemble gets accumulated in the amplitude index $j$ of the auxiliary state, i.e., $\ket{\Phi_{0j}^d}$. Here however we do not measure the auxiliary state to learn the value of $j$. In fact, as shown later, the auxiliary state is kept unchanged (and the entanglement is not consumed).

\begin{figure}
    \centering
    \includegraphics[width=\columnwidth]{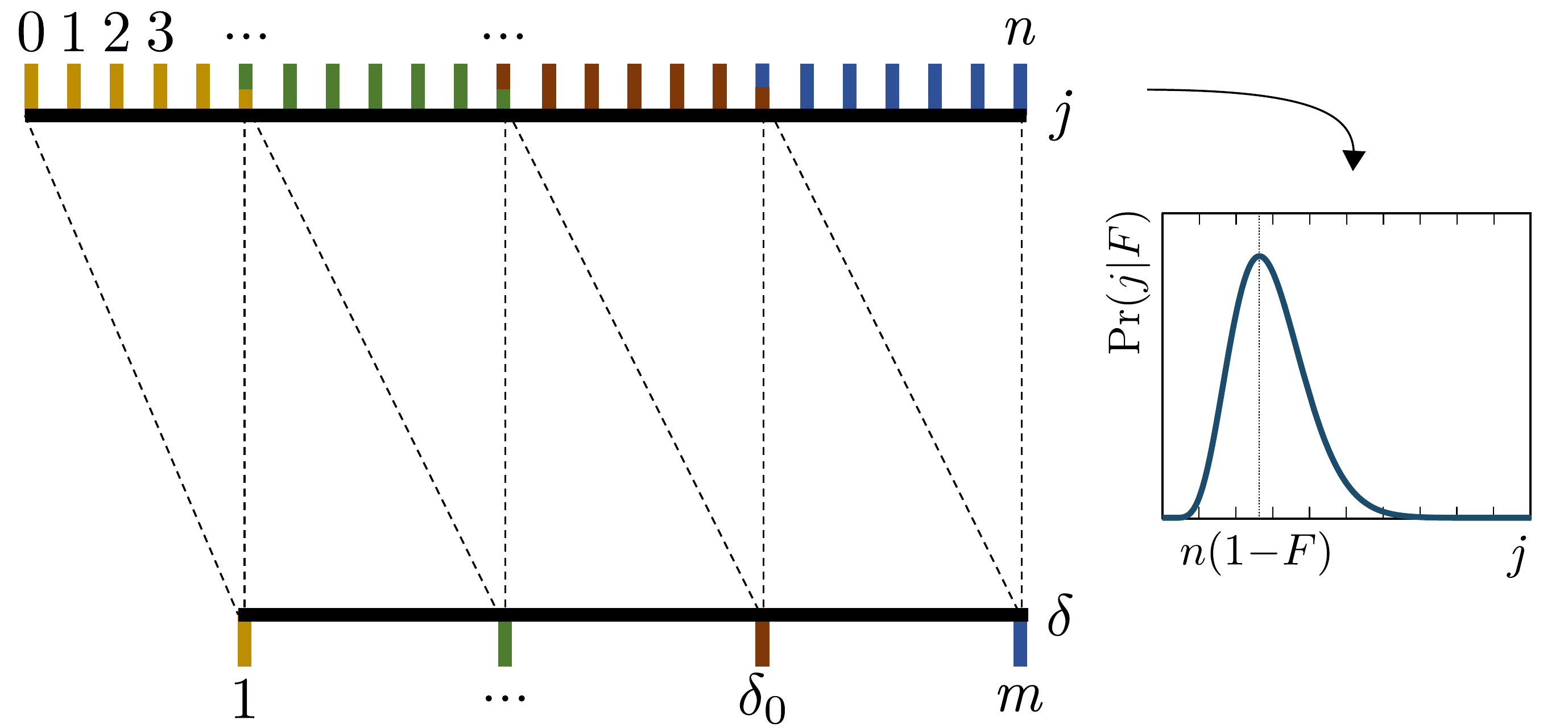}
    \caption{Schematic representation of the coarse-graining protocol P2. Given all the possible values that the amplitude index $j$ of the auxiliary state (upper part) can take, we group them into a few sets encoded in an extra register (lower part). By suitably measuring this extra register, we can learn, up to some failure probability, the information about whether $j$ is above a certain threshold $\delta_0$ without consuming the entanglement of the auxiliary state, whereas the values of $j$ and $\delta$ are directly related with the fidelity $F$ of the states.}
    \label{fig:coarse}
\end{figure}

\begin{figure}
    \subfloat[\centering]{\includegraphics[width=0.5\columnwidth]{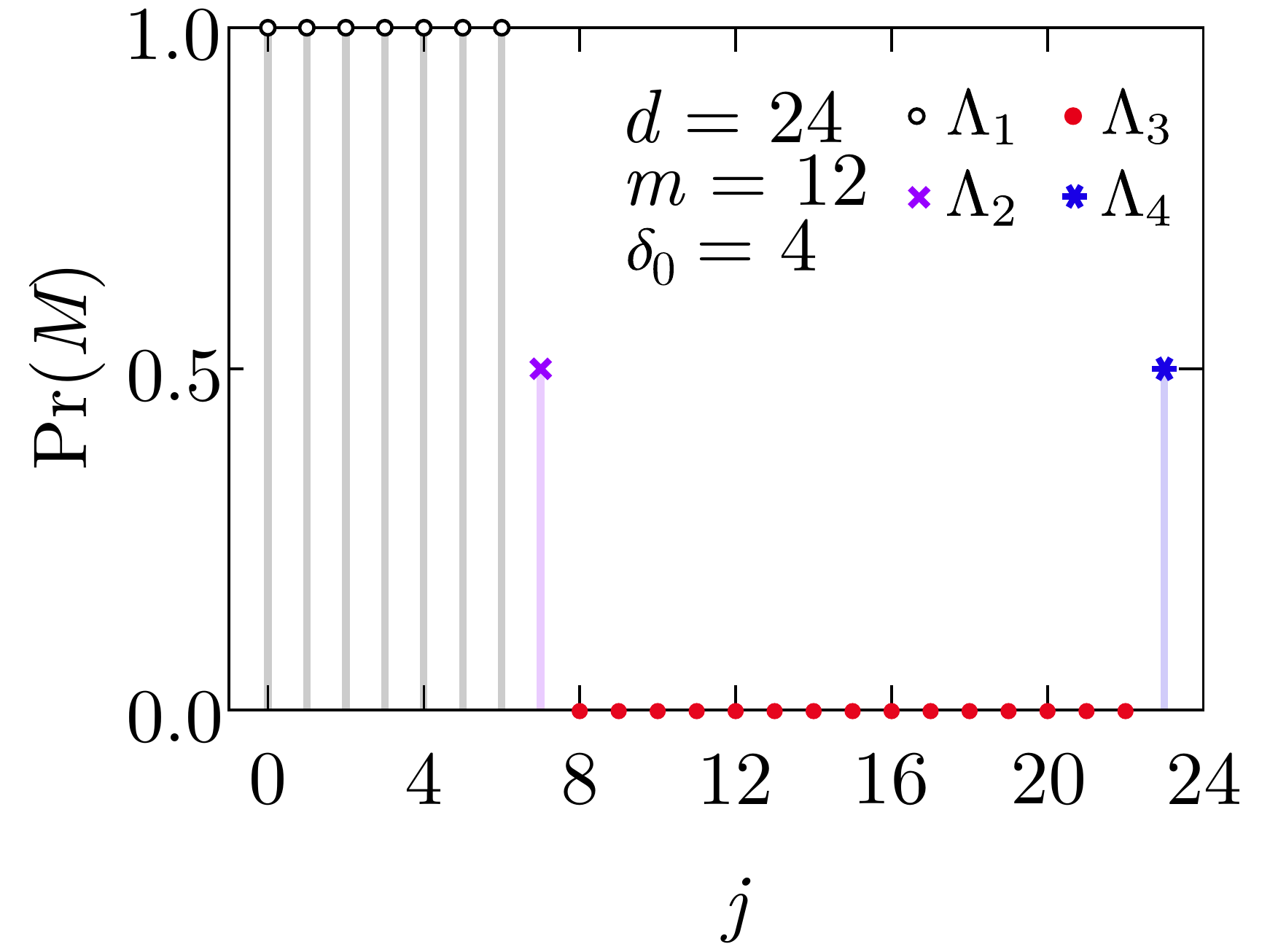} \label{fig:toy:model1} }
    \subfloat[\centering]{\includegraphics[width=0.5\columnwidth]{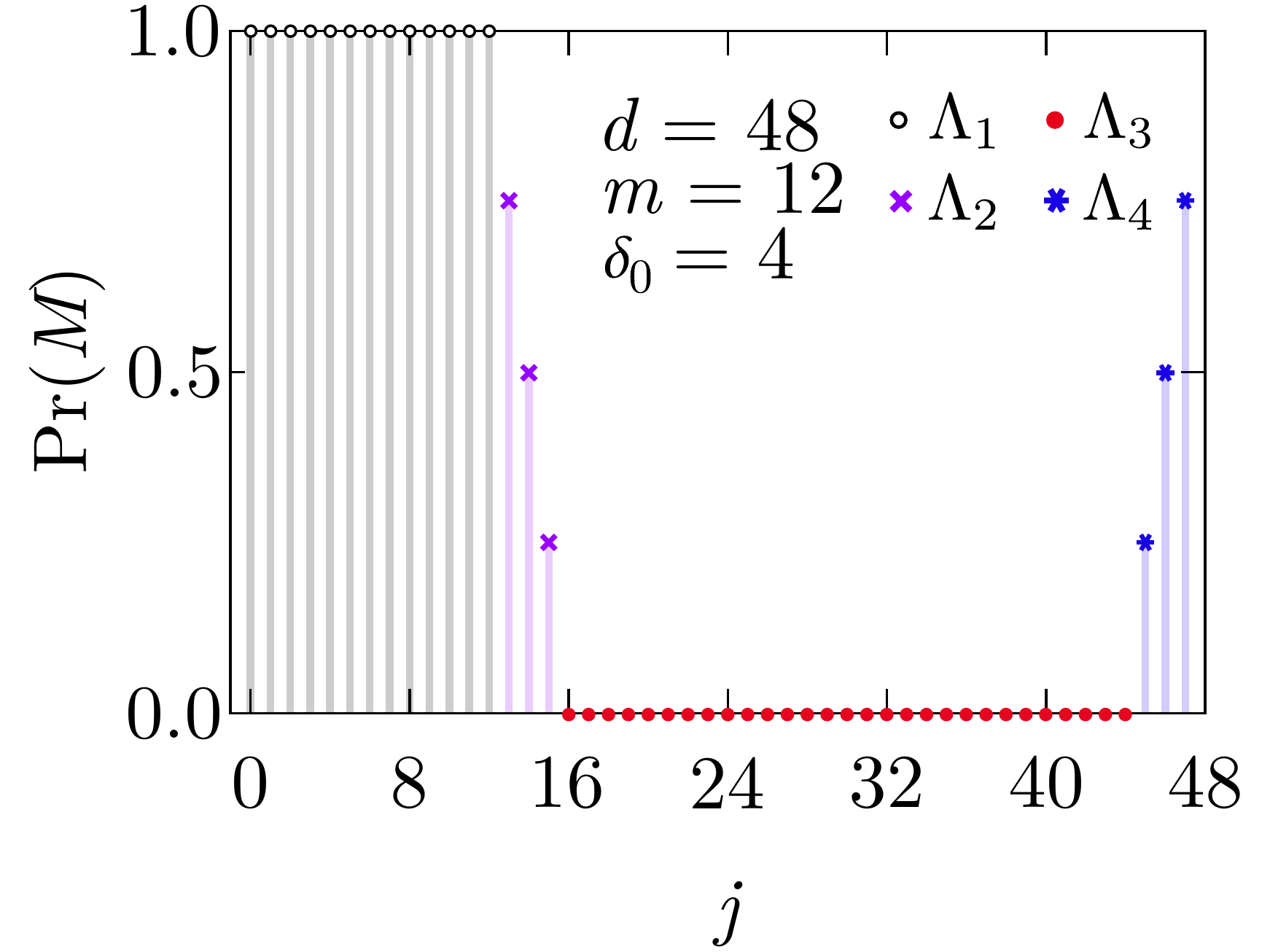} \label{fig:toy:model2} }
    \caption{Analysis of the coarse-graining protocol P2. Given a state $\ket{\Phi^d_{0j}}$, the probability of measuring outcome $M$ as a function of $j$ is plotted for a dimension of the auxiliary state of $d=24$ (a) and $d=48$ (b). The elements of the different sets $\Lambda_i$ are indicated.}
    \label{fig:my_label}
\end{figure}

\begin{table}[h]
    {\LinesNumberedHidden
    \label{table:P2}
    \begin{algorithm}[H]
        \SetKwInOut{Input}{Input}
        \SetKwInOut{Output}{Output}
        \SetAlgorithmName{Protocol P$\!\!$}{}
        \justifying \textit{Input}: Ensemble of $n$ identical noisy Bell states and auxiliary $n+1$-level maximally entangled state.
        \begin{enumerate}
            \item Apply the ENG between the ensemble and the auxiliary state. The information of the noise is accumulated in its amplitude index $j$.
            \item Apply the coarse-grained operation Eq.~\eqref{eq:coarse} from the auxiliary into an additional extra register.
            \item Measure the extra register to learn information about where $j$ lies.
            \item Apply the decorrelation process to recover the auxiliary state untouched. 
        \end{enumerate}
    \justifying \textit{Output}: The witnessing decision problem is solved, i.e., the fidelity of the ensemble states is determined to be above or below some threshold up to some failure probability.
    \caption{coarse-graining}
    \end{algorithm}}
\end{table}

\begin{figure}
    \centering
    \subfloat[\centering]{\includegraphics[width=0.5\columnwidth]{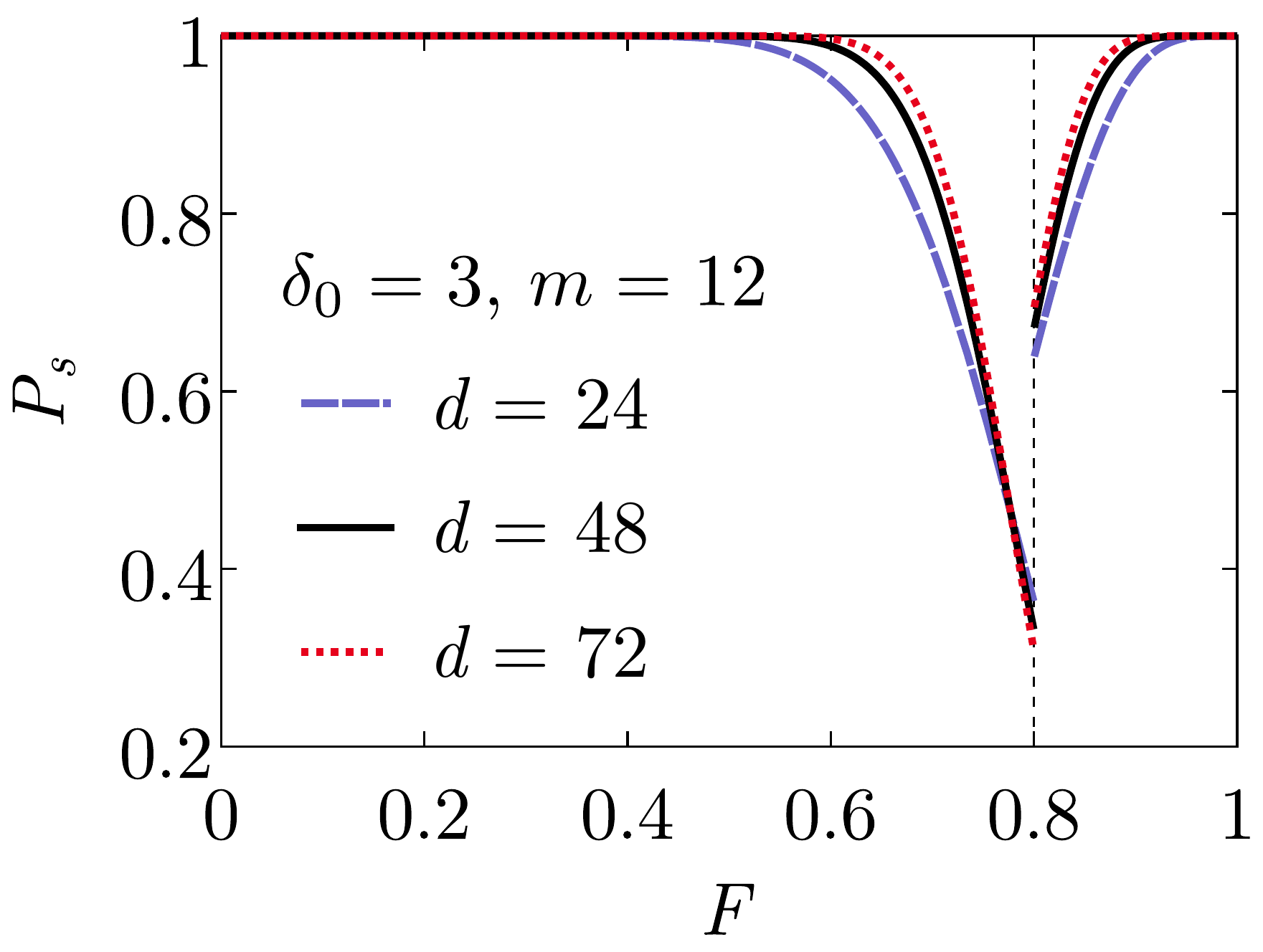} \label{fig:Ps_d} } \subfloat[\centering]{\includegraphics[width=0.5\columnwidth]{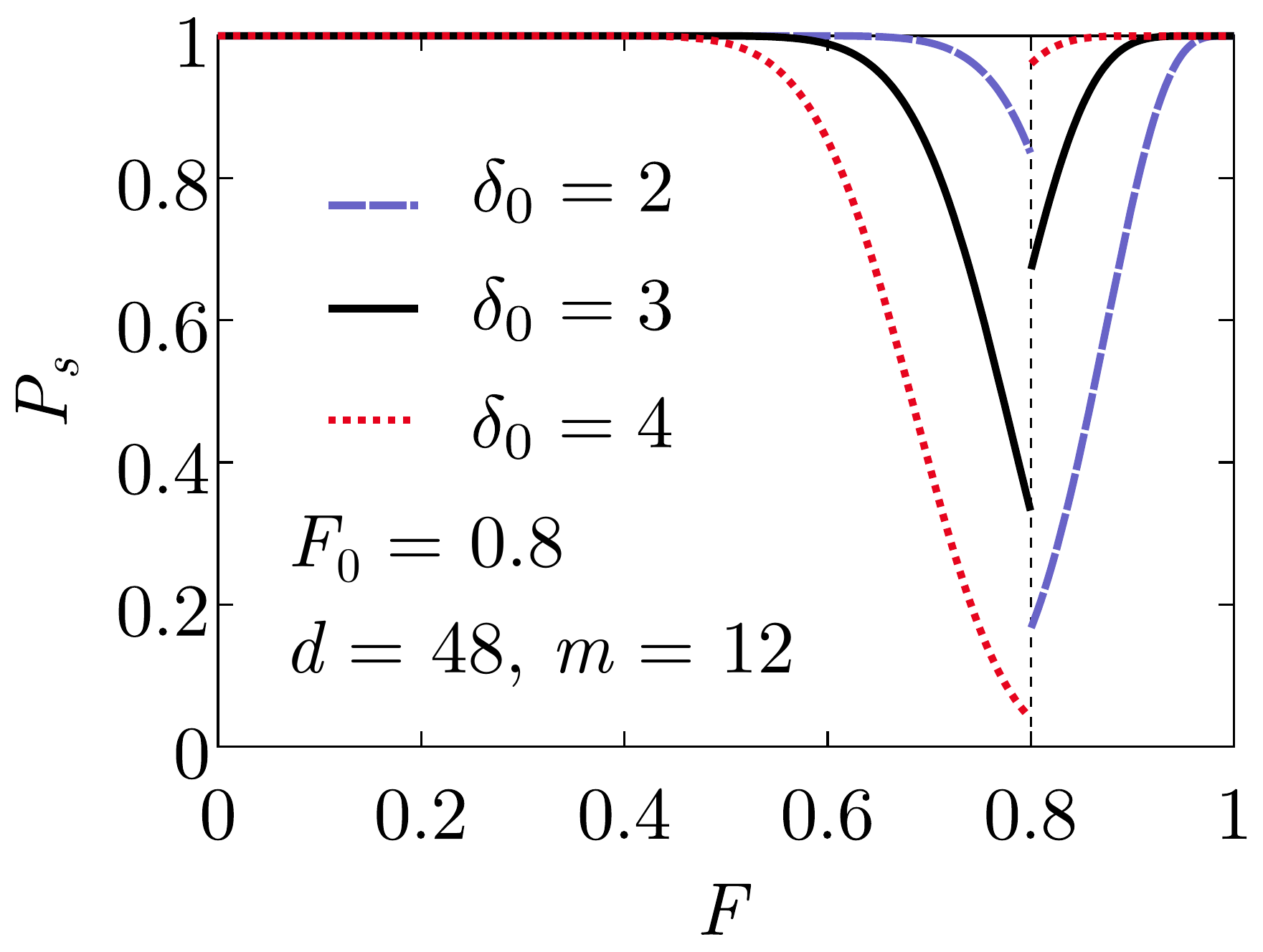} \label{fig:Ps_delta} } \\
    \subfloat[\centering]{\includegraphics[width=0.5\columnwidth]{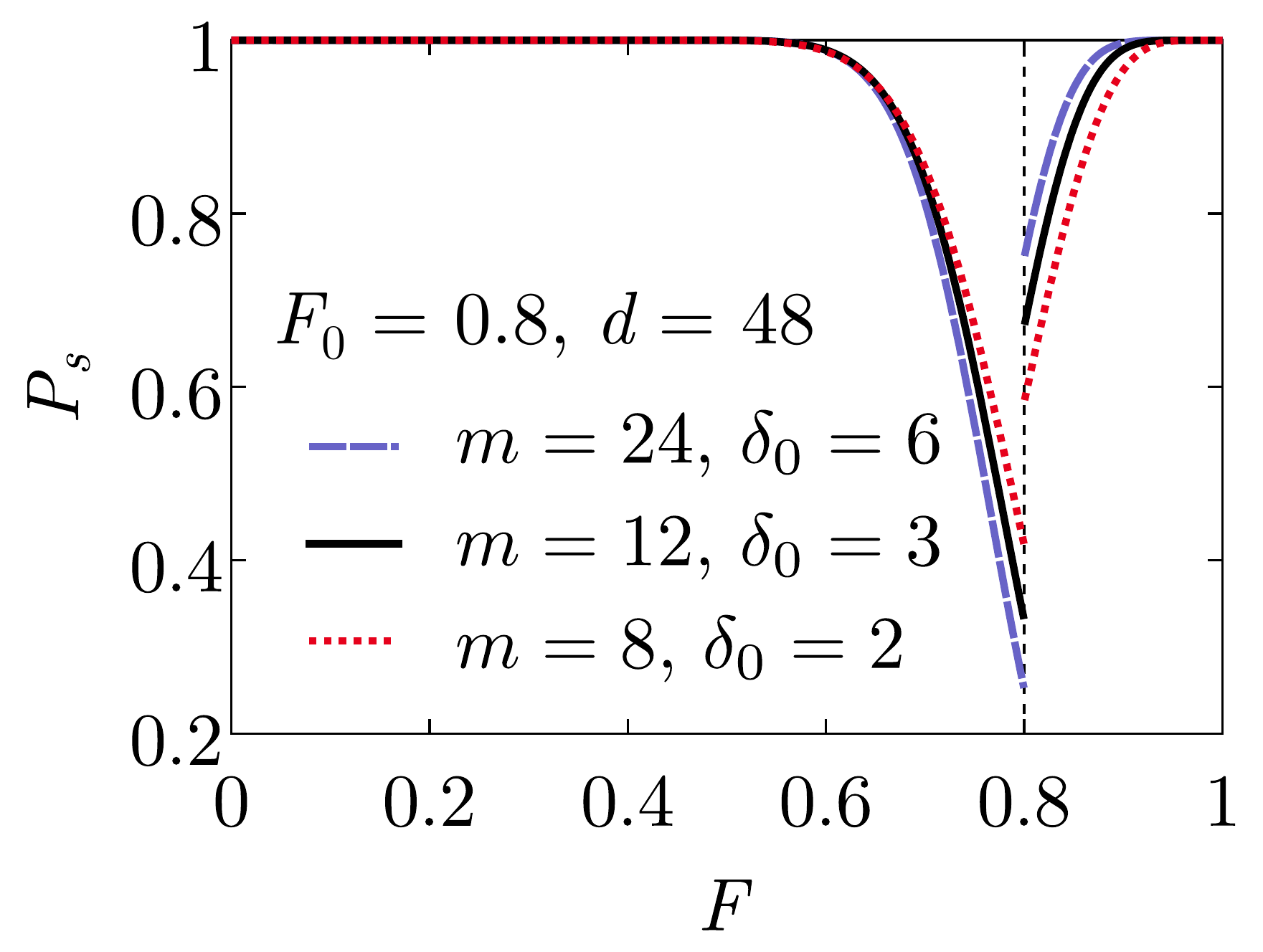} \label{fig:Ps_m} }
    \subfloat[\centering]{\includegraphics[width=0.5\columnwidth]{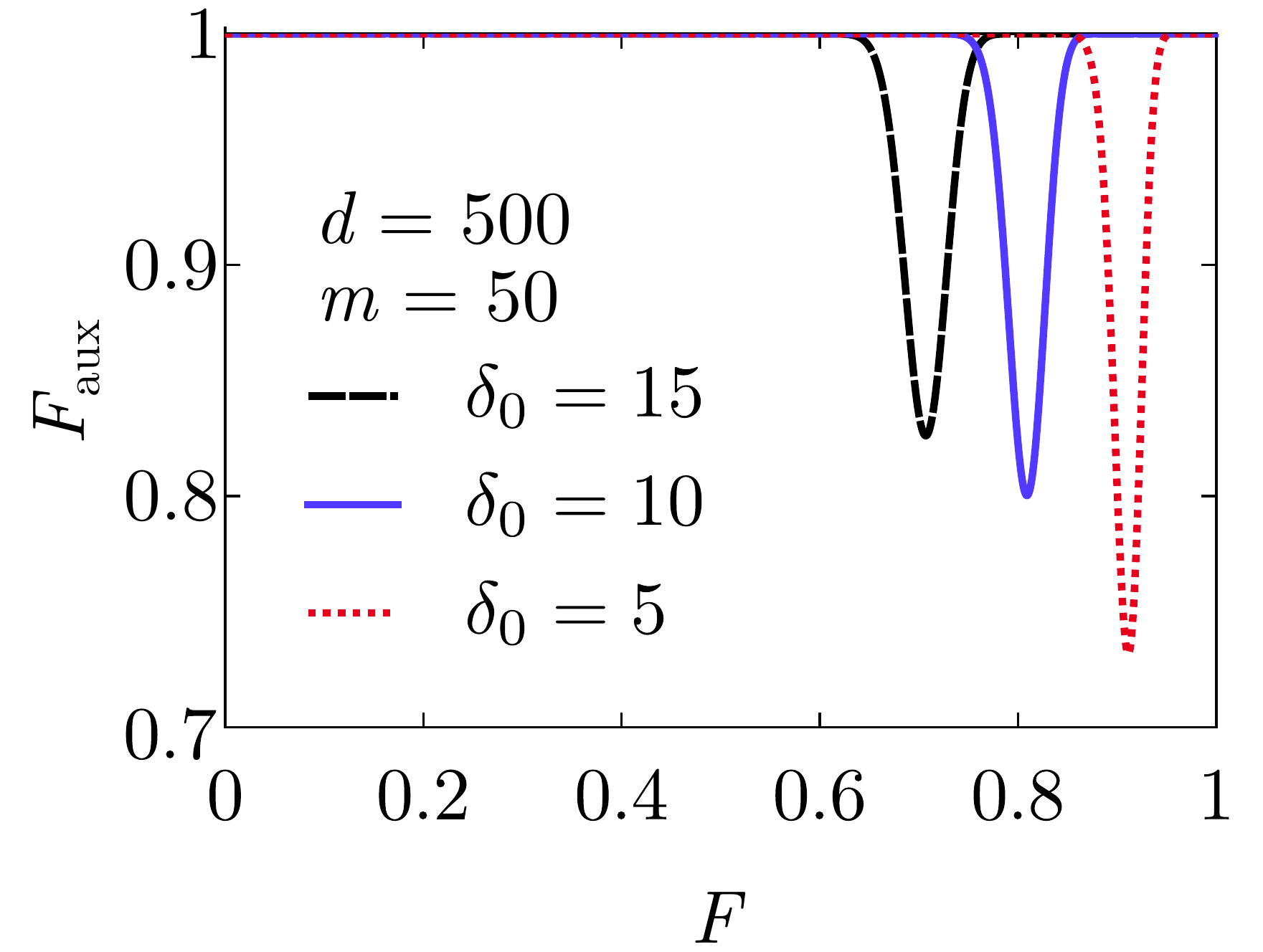} \label{fig:Faux} }
    \caption{Performance of the coarse-graining protocol P2. (a) The success probability of the protocol for a fixed amount of resources consumed (i.e., fixed extra register of dimension $m=12$) for increasing dimension of the auxiliary register $d$. Note that as long as the entanglement of the auxiliary is recovered, the performance of the protocol can be freely enhanced. (b) Success probability for fixed auxiliary states $d=48$ and $m=12$, and decreasing $\delta_0$. (c) Success probability for a fixed auxiliary $d=48$ and with different dimensions of the register $m$. For each value of $m$ we use the maximum ensemble size $n$ and the value of $\delta_0$ that minimizes the discontinuity in $F_0$. (d) Fidelity of the remaining auxiliary state after the implementation of the coarse-gaining protocol P2 for different values of $\delta_0$.}
    \label{fig:coarsegperformance}
\end{figure}

We introduce an additional two-system register $A_{2}B_{2}$, that we denote as \textit{extra register}, of dimension $m \le d$. For simplicity, we assume that $m$ divides $d$, i.e., $d/m \in \mathbbm{N}$, however, this assumption can be lifted. The extra register is initially prepared in the $\ket{00}_{A_2 B_2}$ state locally by parties $A$ and $B$. A bilateral local operation $U_{1 2}$ is applied from the auxiliary to the extra register such that
\begin{equation}
    \label{eq:coarseeq}
    U_{12} = \sum_{k=0}^{d-1} \proj{k}_1 \otimes X^{\left\lceil \frac{k}{d/m}\right\rceil}_2,
\end{equation}
which transfers information of the amplitude index into the extra register in a coarse-graining way (see Fig.~\ref{fig:coarse}). The effect of this operation reads as
\begin{equation}
\label{eq:coarse}
\begin{aligned}
    U_{A_1 A_2} \otimes U_{B_1 B_2} \ket{\Phi^d_{0j}}_{A_1 B_1} \ket{00}_{A_2 B_2} \\
    = \! \frac{1}{\sqrt{d}} \sum_{k=0}^{d-1} \ket{k, \left\lceil \frac{k}{d/m} \right\rceil}_{ \!\! A_1 A_2} \! \ket{k\oplus j,\left\lceil \frac{k \oplus j}{d/m} \right\rceil }_{ \!\! B_1 B_2} \!\!\!.
\end{aligned}
\end{equation}
The different $j$ values get grouped into different sets of certain size and the group or interval $\delta_{0}$ is identified as the interval containing the threshold $j_{0}$ value, where $j_{0}$ is the most likely value of the distribution $\text{Pr}(j|F_{0})$.

The qubit of the extra register that belongs to party $A$ is subsequently teleported to party $B$ --by consuming $\log_2 m$ ebits of entanglement--, such that the global state becomes
\begin{equation}
\label{eq:coarse2}
\begin{aligned}
    \ket{\psi}_{A_1 B_1 B_2 B_3} = \frac{1}{\sqrt{d}} \sum_{k=0}^{d-1} \ket{k}_{A_1} \quad & \\ 
    \otimes \ket{ k \oplus j, \left\lceil \frac{k\oplus j}{d/m} \right\rceil }_{\!\! B_1 B_2} & \ket{\left\lceil \frac{k}{d/m} \right\rceil}_{\!\! B_3},
\end{aligned}
\end{equation}
where we have relabelled qubit $A_2 \rightarrow B_3$. Finally, party $B$ performs a two-outcome projective-operator-valued measure (POVM) measurement defined by the projectors $\{ M, \id - M \}$ on the extra register qubits $B_2, B_3$, i.e.,
\begin{equation}
    \label{eq:measurementextra}
    M = \sum_{\delta = 0}^{\delta_0 - 1} \sum_{l=0}^{m-1} \proj{ \, l \oplus \delta ,l },
\end{equation}
where the projector acts on the $\mathcal{H}_{B_2} \otimes \mathcal{H}_{B_3}$ Hilbert space. 

In case the outcome $M (\id-M)$ is found, we conclude that the fidelity of the probe ensemble states is above (below) the threshold $F_0$, up to a certain failure probability. Importantly, the process only consumes the entanglement required to teleport qubit $A_2$ to party $B$. We consider the number of ebits required by the process (see Sec.~\ref{sec:entanglement:cost}) to evaluate the protocol performance. From a practical perspective, the entanglement required for the teleportation can be obtained by entanglement distillation means (see, e.g., \cite{Riera1, Riera2}) of the ensemble copies directly.

In order to properly understand the effect of the measurement $\{ M, \id - M \}$, we can classify the values of $j$ in four sets (see Fig.~\ref{fig:my_label}): 
\begin{equation*}
\begin{aligned}
    \Lambda_1 & = \big\{ 0, \dots, \, \tilde{d} \, (\delta_0 -1) \, \big\} \\
    \Lambda_2 & = \big\{ \tilde{d} \, (\delta_0 -1) + 1, \dots, \, \delta_0 \tilde{d} -1 \, \big\} \\
    \Lambda_3 & = \big\{ \delta_0 \tilde{d}, \dots, \, \tilde{d} \, (m-1) \, \big\} \\
    \Lambda_4 & = \big\{ \tilde{d} \, (m-1) +1 , \dots, \, d-1 \big\}, 
\end{aligned}
\end{equation*}
where $\tilde{d} = d/m$. The measurement given by Eq.~\eqref{eq:measurementextra} can deterministically distinguish between sets $\Lambda_1$ and $\Lambda_3$, such that in case $j$ lies either in $\Lambda_1$ or $\Lambda_3$, the decision problem is solved deterministically and the auxiliary state is recovered untouched (see below). This is however not the case for the set $\Lambda_2$, for which certain overlapping exists that has to be taken into account. In summary, depending on where the actual set where $j$ lies, the success probability of the witnessing problem reads
\begin{equation*}
\begin{cases}
    \text{Pr} (M) = 1    & \text{ if } \, j \in \Lambda_1                \\
    \text{Pr} (M) = 0    & \text{ if } \, j \in \Lambda_3                \\
    0 < \text{Pr}(M) < 1 & \text{ if } \, j \in \Lambda_2 \cup \Lambda_4.
\end{cases}
\end{equation*}
The performance of the protocol can be directly analyzed given the previous reasoning and given the probability distribution of the $j$ value as a function of the state fidelity. Figure~\ref{fig:coarsegperformance} shows the efficiency for different settings and ensemble sizes. Observe how for the witnessing problem of determining whether the value of the fidelity is $F > F_0 + \frac{\lambda}{2}$ or $F < F_0 - \frac{\lambda}{2}$, the success probability of the protocol can approach $1$ for relatively small additive error $\lambda$.

Once the decision problem is solved, the last step of the protocol entails the recuperation of the auxiliary state, which has been entangled with the extra register during the process. The following steps are required to disentangle the extra register and leave the auxiliary state, and the entanglement associated, untouched. First, the operation $U_{B_1 B_2}$, [Eq.~\eqref{eq:coarseeq}], is undone on the remaining measured state Eq.~\eqref{eq:coarse2}, i.e.,
\begin{equation*}
\begin{aligned}
    U_{B_1 B_2}^\dagger \ket{\psi}_{A_1 B_1 B_2 B_3} & \\ 
    = \frac{1}{\sqrt{d}} \sum_{k=0}^{d-1} \ket{k, k \oplus j, 0, \left\lceil \frac{k}{d/m} \right\rceil }_{\!\! A_1 B_1 B_2 B_3} &.
\end{aligned}
\end{equation*}
Then, qubit $B_3$ is measured in the generalized Fourier basis, given by the basis elements $\{\ket{\alpha_l } = \sum_q \text{exp} (-2 \pi i q/m ) \ket{q} \}$, leading to a state 
\begin{equation*}
\begin{aligned}
    \frac{1}{\sqrt{d}} \sum_{k=0}^{d-1} e^{2 \pi i \left\lceil \frac{k}{d/m} \right\rceil l/m} \ket{k, k\oplus j, 0, {\alpha_l } }_{\! A_1 B_1 B_2 B_3},
\end{aligned}
\end{equation*}
where $l$ refers to the outcome $\ket{\alpha_l}$ obtained. By simply applying a phase gate on party $A$ of the form
\begin{equation*}
    U_{A_1} = \sum_{k=0}^{d-1} e^{ -2 \pi i \left\lceil \frac{k}{d/m} \right\rceil l/m} \proj{k},
\end{equation*}
the initial auxiliary state is recovered.

\textit{Fidelity discrimination.} The coarse-graining strategy is also directly applicable to fidelity discrimination. In this case, to enhance its performance, the dimension of the extra register should be chosen depending on the possible fidelity values $F_1$, $F_2$. Performance is comparable to the efficiency of the witnessing problem.

\subsection{Protocol P3: Ensemble blocking }

We finally consider a strategy based on parity measurements of subsets or blocks of the ensemble. In contrast to the previously discussed protocols P1 and P2, this approach is not limited to states $\rho_a$ resulting from amplitude damping, but also to other state families such as Werner states $\rho_{\text{w}}$.  

\begin{figure}
    \centering
    \includegraphics[width=\columnwidth]{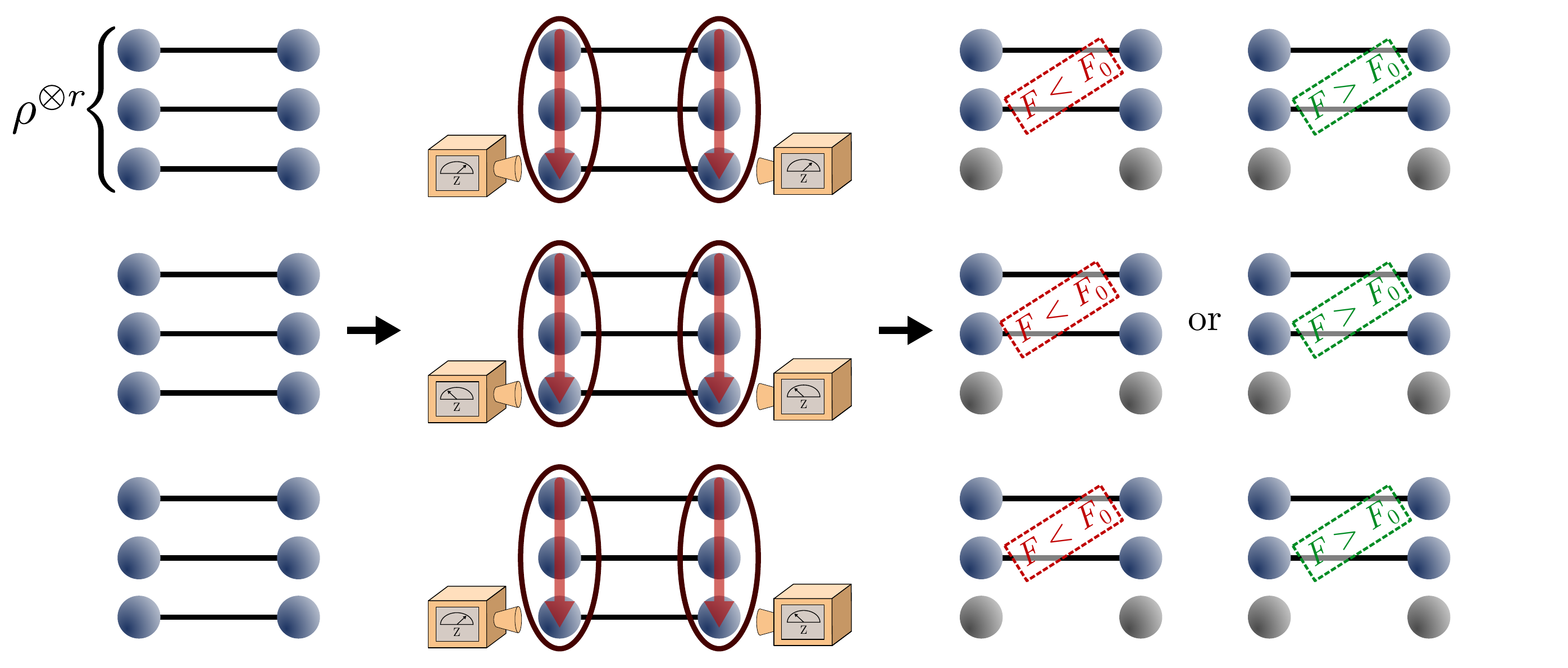} 
    \caption{\label{fig:blockingscheme} Schematic representation of the blocking strategy (P3). The initial ensemble of $n$ Bell diagonal states is divided into several blocks of size $r$. Then, the parity of each block is encoded in one of the states which later is measured revealing the actual parity value of the block. From the number of obtained ``even'' parties, we can determine if the initial fidelity $F$ was above or below the threshold $F_0$ with some success probability.}
\end{figure}

\textit{Fidelity witnessing.} Given an ensemble of $N$ unknown states we first depolarize them into a Werner state form, i.e., $\rho^{\otimes N} \mapsto \rho_{\text{w}}^{\otimes N}$, see Sec.~\ref{sec:depolarization}. Then, we divide the ensemble into $n$ blocks of $r$ states each. Making use of bilateral CNOT gates, in analogy to purification hashing techniques \cite{Bennett_hashing}, which act locally from the states of each block into one of the states, we can learn the parity of each block (see Fig.~\ref{fig:blockingscheme}), i.e., 
\begin{equation}
\label{eq:bCNOT}
\begin{split}
    \text{CNOT}^{A_1 A_2}_{1 \rightarrow 2} \otimes \text{CNOT}^{B_1 B_2}_{1 \rightarrow 2} |\Psi_{ij}\rangle_{A_1 B_1} |\Psi_{kl} \rangle_{A_2 B_2} & \\ = \, |\Psi_{i\oplus k, j} \rangle_{A_1 B_1}|\Psi_{k, l\oplus j} \rangle_{A_2 B_2} &.
\end{split}
\end{equation}
Note that maximally entangled copies are not required since the parity can be encoded in one of the states of the block. We denote as $\kappa$ the number of \textit{even} parties obtained, whose probability follows a binomial distribution of the form
\begin{equation*}
    \text{Pr}(\kappa|\rho) = \binom{n}{\kappa} \big[ \pi_0 \! \left( \rho^{\otimes r} \right) \big]^{n-\kappa} \big[ 1 - \pi_0 \! \left( \rho^{\otimes r} \right) \big]^{\kappa},
\end{equation*}
where the probability of measuring even parity in a block of size $r$ is
\begin{equation*}
    \pi_0 \! \left(\rho^{\otimes r}\right) = \sum_{k=0}^{ r/2 } \binom{r}{2k} \, A^{r-2k} \left( 1-A \right)^{2k},
\end{equation*}
where $A \equiv (1+2F)/3$.

Since $\kappa$ is given by a binomial distribution, we can then repeat the same analysis performed for Protocol P0 (see Sec.~\ref{sec:protocol0}), but using $\text{Pr}(\kappa|\rho)$ instead of $\text{Pr}(j|F)$. Observe how, despite $N$ states being involved in the protocol, only $n$ are destroyed and the remaining $(r-1)n$ states are now correlated. After the first parity round is complete, the fidelity of the remaining states is changed depending on the value of the parity obtained. Due to the back-action effect of the bilateral control gate, see Eq.~\eqref{eq:bCNOT}, the expected value of the local fidelity is given by
\begin{equation*}
    F' = F^2 + F \left(\frac{1-F}{3}\right) + 2\left(\frac{1-F}{3} \right)^2.
\end{equation*}

The performance of the protocol can be further enhanced in case the initial states of the ensembles are of a certain form. One case of particular interest involves states affected by dephasing type noise, Eq.~\eqref{eq:depashing}, for which the first depolarization step is not required. For this kind of state, there is no back-action effect from the bilateral CNOT gate, and hence the average fidelity remains unchanged, i.e., $F' = F$.

\begin{table}[h]
    {\LinesNumberedHidden
    \label{table:P3}
    \begin{algorithm}[H]
        \SetKwInOut{Input}{Input}
        \SetKwInOut{Output}{Output}
        \SetAlgorithmName{Protocol P$\!\!$}{}
        \justifying \textit{Input}: Ensemble of $n$ identical noisy Bell states.
        \begin{enumerate}
            \item Depolarize the ensemble to the form of Eq.~\eqref{eq:werner}.
            \item Apply bilateral CNOT gates Eq.~\eqref{eq:bCNOT} from ensemble blocks to some target ensemble states.
            \item Learn the parity of each block by measuring the target states.
            \item If $\kappa \in \sigma$ assume $F>F_0$, where $\kappa$ is the number of even parities.
        \end{enumerate}
    \justifying \textit{Output}: The fidelity of the initial ensemble was above or below $F_0$ with some success probability $P_s$.
    \caption{Ensemble blocking}
    \end{algorithm}}
\end{table}

Additionally, one can implement a second round of parity measurements to further enhance the performance. This is accomplished by simply learning again the parity of the blocks where an even parity was found, leading to a recursive improvement in the protocol efficiency.

From the value of $\kappa$ we can again solve the fidelity witnessing problem following an analogous analysis as the one described for protocol P0, Sec.~\ref{sec:protocol0}, by including the block size $r$ as an extra parameter to consider, such that the success probability can be optimized by simply defining a block size $r^*$ that maximizes it i.e.,
\begin{equation*}
    \int_0^1 P_s(r^*,F) \, \mathrm{d} F = \max_r \int_0^1 P_s(r,F) \, \mathrm{d} F.
\end{equation*}

\textit{Fidelity discrimination.} The value of even parities $\kappa$ can be also used to solve the discrimination problem. In this case, the optimal $r^*$ is such that the difference of obtaining an even parity is maximum, i.e.,
\begin{equation*}
    \pi_0 \big(\rho_1^{\otimes r^*}\big) - \pi_0 \big(\rho_2^{\otimes r^*}\big) = \max_r \pi_0 \! \left(\rho_1^{\otimes r}\right) - \pi_0 \! \left(\rho_2^{\otimes r}\right),
\end{equation*}
where the larger the difference, the more distinguishable the two probability distributions are. Importantly, the blocking strategy allows us to overcome the optimal success probability for fidelity discrimination by involving the whole ensemble in the process but only partially consuming it.

\begin{figure}
    \centering
    \subfloat[\centering]{\includegraphics[width=0.48\columnwidth]{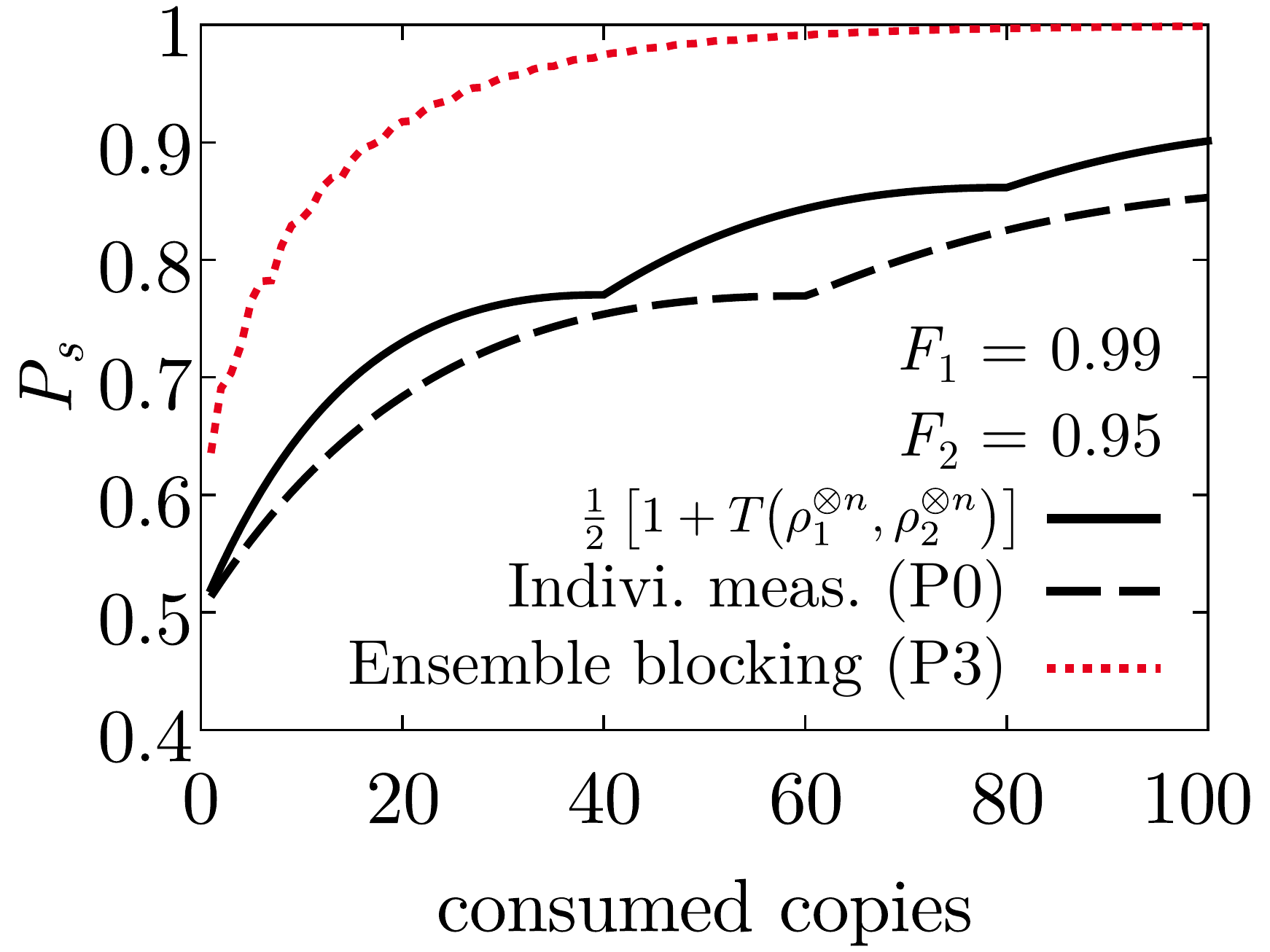} \label{fig:block1} }
    \subfloat[\centering]{\includegraphics[width=0.48\columnwidth]{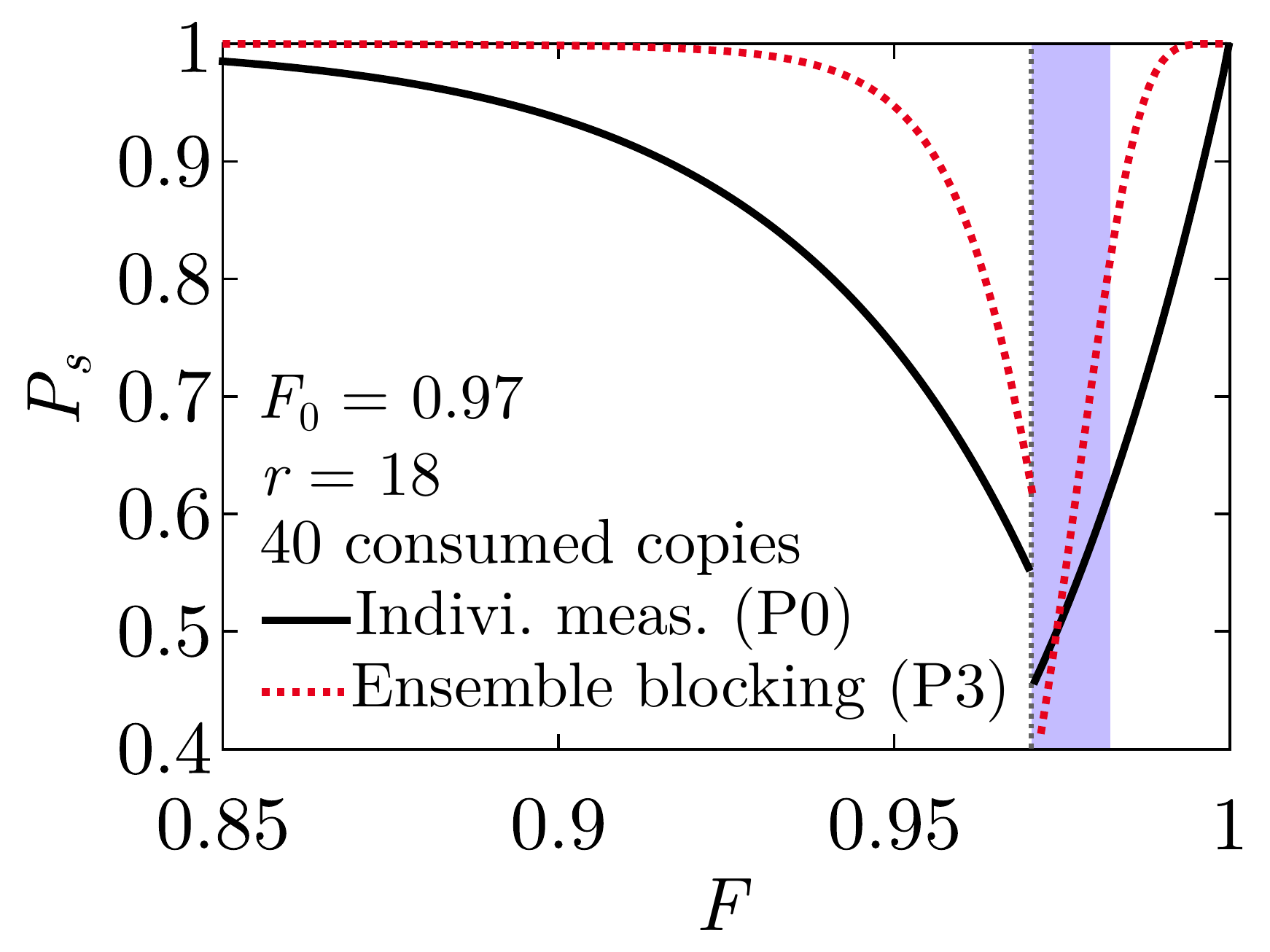} \label{fig:block2} } \hfill
    \subfloat[\centering]{\includegraphics[width=0.48\columnwidth]{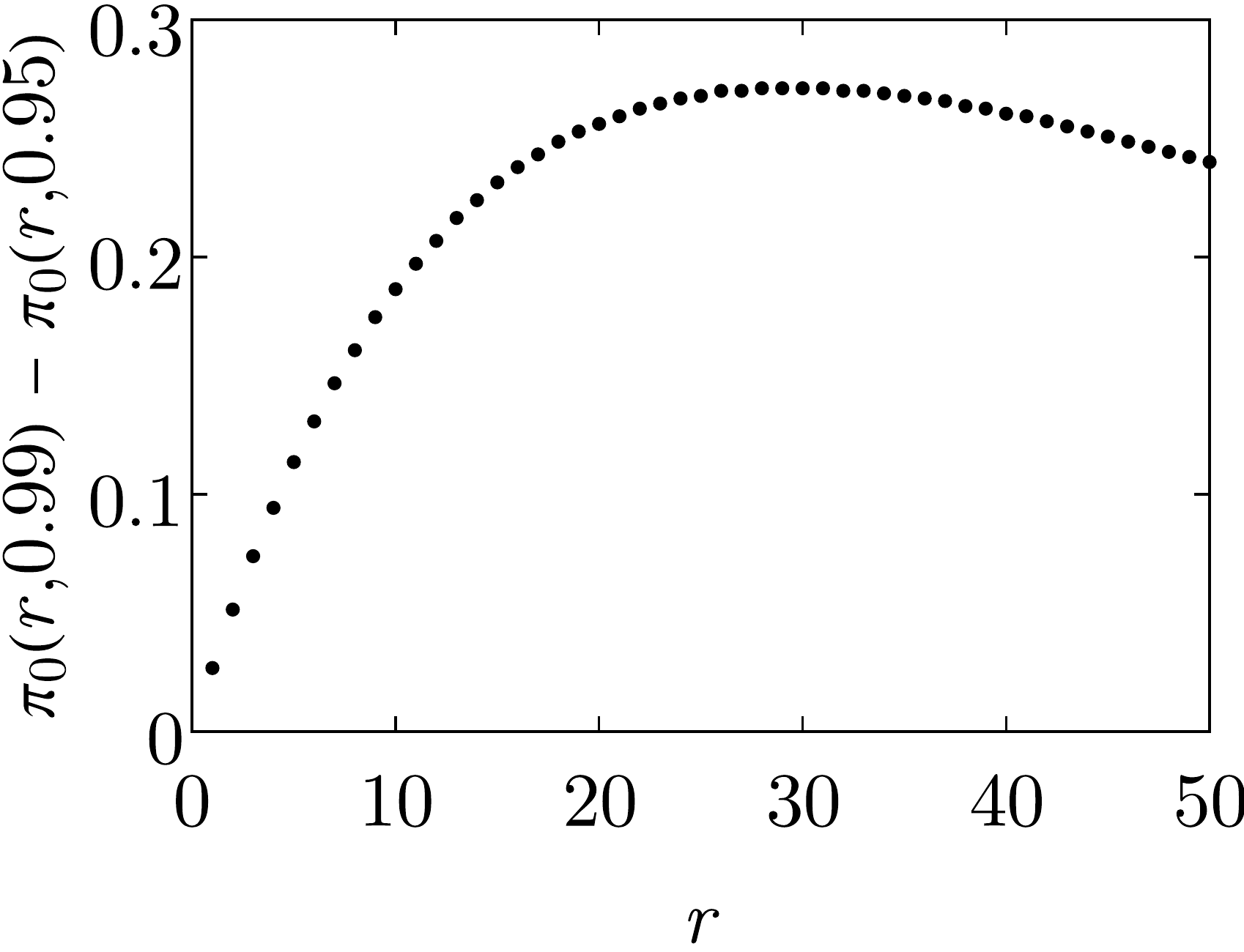} \label{fig:block3} }
    \subfloat[\centering]{\includegraphics[width=0.48\columnwidth]{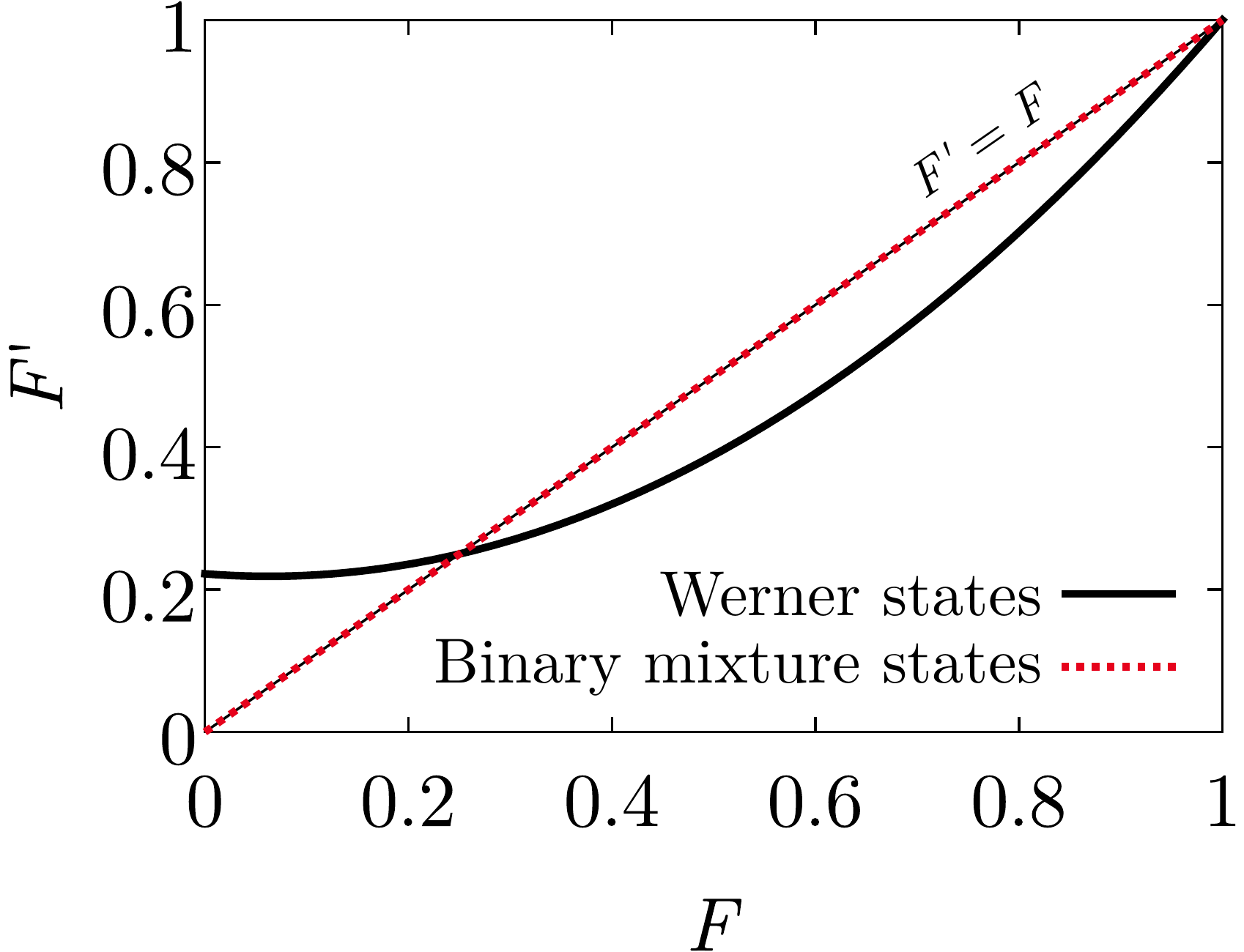} \label{fig:block4} }
    \caption{\label{fig:bloking} Performance of the blocking protocol for Werner-type states. (a) Fidelity discrimination with equal prior probability for each state, i.e., $\eta_1 = \eta_2 = 1/2$. Success probability as a function of the number of consumed copies. (b) Fidelity witnessing with $F_0 = 0.97$. Success probability as a function of the fidelity of the ensemble. The shadow region corresponds to the range of $F>F_0$ such that $F'<F_0$. (c) Difference of probability of obtaining parity 0 as a function of the block size $r$ for two blocks of fidelities $F_1 = 0.99$ and $F_2=0.95$. The difference is maximized for $r=29$. (d) Fidelity of the remaining states of the ensemble after the application of a singer round of the blocking protocol.}
\end{figure}

Figures \ref{fig:block1} and \ref{fig:block2} show the performance of the blocking strategy, compared with protocol P0 based on individual measurements. Observe how the blocking strategy provides significant performance enhancement with respect to the individual measurements strategy (P0).

\section{Comparison of approaches}

In order to compare different protocols, we use the number of required resources $R$ to obtain a certain success probability $P_s$ as a measure. We demand that the success probability for a fixed fidelity is the same for all protocols. We use the success probability of the single-copy approach P0 as a reference value, and adjust the parameters of the other protocols to guarantee that they lead to the same or a larger success probability. The resources are the number of copies that are measured and hence destroyed. Notice that for the error counting and coarse-graining protocols, P1 and P2, we also take the cost for used maximally entangled auxiliary states into account. In order to do so, we assume that these auxiliary states are generated from noisy states of the ensemble, e.g., via entanglement purification, where the yield ($Y$) of the entanglement purification protocol (a combination of the recurrence protocol of \cite{Deutsch1996} and hashing \cite{Bennett_hashing}) determines how many noisy copies correspond to a perfect ebit. For an entangled auxiliary register of dimension $d$, $\log_2(d)/Y$ states of the ensemble are required.

An overview of the applicability and strengths of the different protocols is given in Table~\ref{tab:1}. All protocols are applicable for states of the form $\rho_a$ resulting from amplitude damping, where both P1 and P2 give an (up-to) exponential advantage as compared to the standard approach P0. Notice though that the error counting and coarse-graining protocols P1 and P2 are not efficient for Werner states. However, the blocking protocol P3 still outperforms the standard approach P0 also in this case, as is shown in Figs. \ref{fig:block1} and \ref{fig:block1}.

\begin{figure}
\label{Fig:comparison}
    \centering
    \includegraphics[width=\columnwidth]{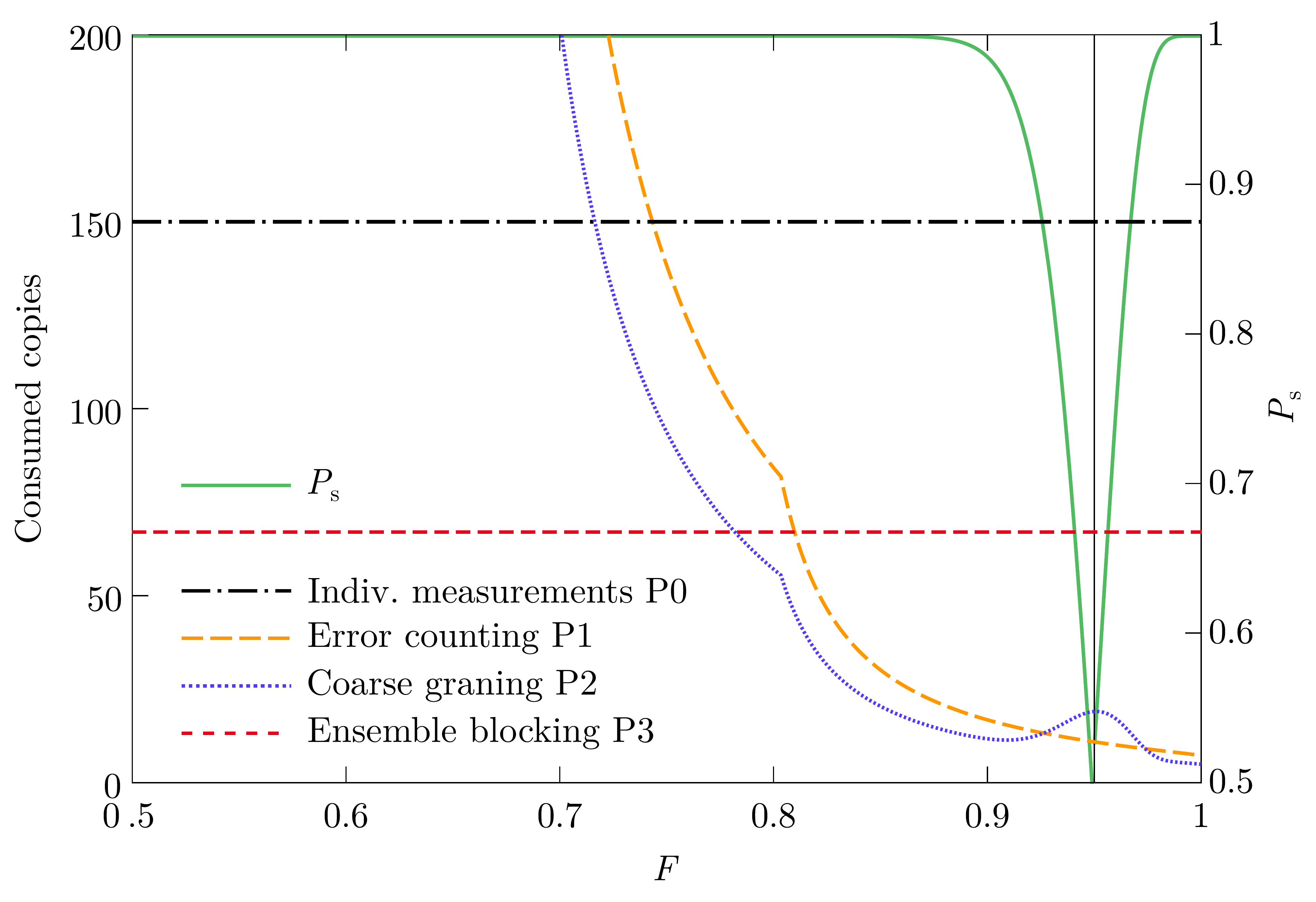}
    \caption{\label{fig:comparison} Comparison between different protocols. The required resources $R$, i.e., the number of consumed copies, are plotted as a function of the fidelity for different protocols. We assume a threshold fidelity of $F_0=0.95$, and an ensemble of size $n$ corresponding to states resulting from amplitude damping, $\rho_a$. We consider a fixed success probability for all protocols that varies with the fidelity (right vertical axis). Notice that ensemble sizes differ for different protocols in order to match success probabilities, but the relevant quantity is the consumed resources which are plotted. Parameter used for the different protocols are as follows; P0: ensemble size $n=150$, where all copies are measured; P1: ensemble size $n=150$, dimension $d= \log_2 151$ for auxiliary register; P2: ensemble size $n=290$, dimension of auxiliary register $d=300$, dimension of coarse-grained register $m=30$, $\delta_0 = 2$, P3: ensemble size $n=603$, block size $r=9$ measured copies $67$.}
\end{figure}

In Fig.~\ref{Fig:comparison} we compare the different protocols (see also Table~\ref{tab:1}), and plot the required resources $r$ as a function of the fidelity of the initial states in the ensemble for the fidelity witnessing problem, i.e., to decide if the state has a fidelity larger or smaller than $F_0$. One clearly observes that P1 and P2 offer a large improvement for fidelities that are close to the threshold fidelity $F_0$, though they work even better if one excludes a small interval around $F_0$, i.e., considers the promise problem that $F \geq F_0 + \lambda/2$ or $F< F_0 - \lambda/2$. Similarly, for the fidelity discrimination problem, one obtains an even larger improvement for P1 and P2.

\renewcommand{\arraystretch}{1.4}
\begin{table*}[]
\label{Table:comparison}
\begin{tabular}{c|c|c|c|c|}
\cline{2-5}
 & \multicolumn{4}{c|}{Protocol efficiency: required resources $R$ } \\ \hline
\multicolumn{1}{|c|}{State type} & $\,$Indiv. meas. (P0)$\,$   & $\,$Error counting (P1)$\,$                               & $\,$Coarse-graining (P2)$\,$  & $\,$Ensemble blocking (P3)$\,$                 \\ \hline \hline
\multicolumn{1}{|c|}{$\,$Amplitude damping $\rho_{a}\,$}     & \multirow{3}{*}{$R_0$} & $\mathcal{O}\left[ \log{ R_0 } \right]$ & $\mathcal{O}\left[\alpha (n) \, \log{ R_0 } \right]$ & \multirow{3}{*}{$\mathcal{O}\left[\beta(F_0) \, R_0\right]$} \\ \cline{1-1} \cline{3-4}
\multicolumn{1}{|c|}{Dephasing $\rho_{\text{d}}$}            &                        & \multirow{2}{*}{non-efficient} & \multirow{2}{*}{non-efficient} & \\ \cline{1-1}
\multicolumn{1}{|c|}{Werner $\rho_{\text{w}}$}               &                        &                          &                                      & \\ \hline
\end{tabular}
\caption{\label{tab:1} General comparison between protocols. We use the required resources $R$ to obtain a certain success probability $P_s$ as a figure of merit to assess the performance of the different protocols, where we denote by $R_0$ the required resources of the reference protocol P0 to achieve $P_s$, and we demand the same or larger success probability for the other protocols. The resources depend on the problem setting (threshold fidelity $F_0$ in witnessing and $F_1, F_2$ in discrimination) and the required success probability. $\alpha (n)$ and $\beta(F_0)$ are factors that depends on the ensemble size or the threshold fidelity respectively, and which are typically smaller than 1, i.e., they indicate an improvement over P0. Efficient variants of the P1 and P2 protocols that offer an exponential improvement over P0 are not known for dephasing and Werner-type states.}
\end{table*}

\section{Conclusions and outlook}
\label{sec:conclusions}

In this work, we have considered the verification of noisy entangled states, with the aim to decide if the quality of states in a (large) ensemble is sufficient to use them for some desired application. We introduced methods to distinguish between two sets of entangled states by means of local operations and classical communication, eventually assisted by entanglement. We have concentrated on specific state families and fidelity as the central feature. Specifically, we introduced protocols to solve the decision problems of determining if the ensemble consists of states with fidelity $F_1$ or $F_2$ (discrete sets), or if the fidelity of the states is above or below a certain threshold value $F_0$, possibly excluding a small interval around $F_0$. The nature of the problem that we called fidelity witnessing requires as output only one bit of information, in contrast to well-studied problems such as state tomography or fidelity estimation, where a significantly larger amount of information needs to be determined. As a first result, we have found that this practically relevant decision problem can for some state families, e.g., resulting from imperfect storage with decay as the dominant noise source, be solved more efficiently than using fidelity estimation or full state tomography.

Perhaps more importantly, we demonstrate that using a larger ensemble while measuring and hence consuming only a small subset of states, provides a significant advantage. This is similar in spirit as utilized in \cite{Riera1,miguel2022improving} in the context of entanglement purification or state certification, but generalizes and extends these ideas in a non-trivial way and makes them applicable to new problems. This in fact leads to an up-to-exponential improvement as compared to methods, extensions of state verification \cite{Wang2019, Li19, Zhu1, Zhu2, Zhu19, Yu_2019}, that operate on ensembles of a fixed size where all states are measured. Some protocols we introduce in this context operate on states of the ensemble directly (protocol P3), without any extra resources, where blocks of a certain size are locally manipulated by collective operations and only a few states are measured. The rest of the ensemble remains intact and can be used for the desired application after successful verification. Other approaches we introduce, such as protocols P1 and P2, require auxiliary entangled states to write in and read out relevant information of the whole ensemble. While in protocol P1 this auxiliary register is measured and hence destroyed, in protocol P2 the information is first coarse-grained, and most of the auxiliary entanglement is preserved and can be recovered. Only the read-out of a small amount of coarse grained information is required, and little entanglement is consumed to access this (non-local) information with local operations only. It is in fact the latter protocol that yields a provable exponential advantage as compared to strategies that operate on fixed-size ensembles that are fully measured. Notice that consumed auxiliary entanglement can be directly related to the number of copies of states in the ensemble that need to be measured. One may either actually use noisy states from the ensemble as auxiliary states, or first, produce, e.g., by means of entanglement purification, high-fidelity or even perfect auxiliary entangled states from noisy copies. The conversion rate is given by the performance of entanglement purification, where reachable bounds are known. One can hence translate also protocols using auxiliary entanglement into schemes that only use noisy states from the ensemble, and find the total number of consumed states. This allows one to compare the different strategies, and assess performance (e.g., success probability) for a given number of consumed copies, or the required number of copies to reach a certain accuracy.

We remark that the ideas and tools we present here are not limited to the specific state classes we consider, but may be more broadly applicable. For instance, we believe that a generalization to multipartite entangled states such as Greenberger-HorneZeilinger (GHZ) states or certain graph states is straightforward, and will be presented elsewhere. Also, the idea of using a larger ensemble, concentrating information and measuring only a small fraction might be useful for other related problems, such as, e.g., for improving fidelity estimation.

\vspace{-1cm}
\section*{Acknowledgments}
This work was supported by the Austrian Science Fund (FWF) through
Projects No. P30937-N27, No. P36009-N, and No. P36010-N. Finanziert von der Europ\"aischen Union - NextGenerationEU.

\bibliographystyle{apsrev4-1}
\bibliography{Fidelity_witnessing_biblio.bib}

%merlin.mbs apsrev4-1.bst 2010-07-25 4.21a (PWD, AO, DPC) hacked
%Control: key (0)
%Control: author (72) initials jnrlst
%Control: editor formatted (1) identically to author
%Control: production of article title (-1) disabled
%Control: page (0) single
%Control: year (1) truncated
%Control: production of eprint (0) enabled
\begin{thebibliography}{46}%
\makeatletter
\providecommand \@ifxundefined [1]{%
 \@ifx{#1\undefined}
}%
\providecommand \@ifnum [1]{%
 \ifnum #1\expandafter \@firstoftwo
 \else \expandafter \@secondoftwo
 \fi
}%
\providecommand \@ifx [1]{%
 \ifx #1\expandafter \@firstoftwo
 \else \expandafter \@secondoftwo
 \fi
}%
\providecommand \natexlab [1]{#1}%
\providecommand \enquote  [1]{``#1''}%
\providecommand \bibnamefont  [1]{#1}%
\providecommand \bibfnamefont [1]{#1}%
\providecommand \citenamefont [1]{#1}%
\providecommand \href@noop [0]{\@secondoftwo}%
\providecommand \href [0]{\begingroup \@sanitize@url \@href}%
\providecommand \@href[1]{\@@startlink{#1}\@@href}%
\providecommand \@@href[1]{\endgroup#1\@@endlink}%
\providecommand \@sanitize@url [0]{\catcode `\\12\catcode `\$12\catcode
  `\&12\catcode `\#12\catcode `\^12\catcode `\_12\catcode `\%12\relax}%
\providecommand \@@startlink[1]{}%
\providecommand \@@endlink[0]{}%
\providecommand \url  [0]{\begingroup\@sanitize@url \@url }%
\providecommand \@url [1]{\endgroup\@href {#1}{\urlprefix }}%
\providecommand \urlprefix  [0]{URL }%
\providecommand \Eprint [0]{\href }%
\providecommand \doibase [0]{http://dx.doi.org/}%
\providecommand \selectlanguage [0]{\@gobble}%
\providecommand \bibinfo  [0]{\@secondoftwo}%
\providecommand \bibfield  [0]{\@secondoftwo}%
\providecommand \translation [1]{[#1]}%
\providecommand \BibitemOpen [0]{}%
\providecommand \bibitemStop [0]{}%
\providecommand \bibitemNoStop [0]{.\EOS\space}%
\providecommand \EOS [0]{\spacefactor3000\relax}%
\providecommand \BibitemShut  [1]{\csname bibitem#1\endcsname}%
\let\auto@bib@innerbib\@empty
%</preamble>
\bibitem [{\citenamefont {Ekert}(1991)}]{Ekert91}%
  \BibitemOpen
  \bibfield  {author} {\bibinfo {author} {\bibfnamefont {A.~K.}\ \bibnamefont
  {Ekert}},\ }\href {\doibase 10.1103/PhysRevLett.67.661} {\bibfield  {journal}
  {\bibinfo  {journal} {Phys. Rev. Lett.}\ }\textbf {\bibinfo {volume} {67}},\
  \bibinfo {pages} {661} (\bibinfo {year} {1991})}\BibitemShut {NoStop}%
\bibitem [{\citenamefont {Lo}\ \emph {et~al.}(2014)\citenamefont {Lo},
  \citenamefont {Curty},\ and\ \citenamefont {Tamaki}}]{Lo2014}%
  \BibitemOpen
  \bibfield  {author} {\bibinfo {author} {\bibfnamefont {H.-K.}\ \bibnamefont
  {Lo}}, \bibinfo {author} {\bibfnamefont {M.}~\bibnamefont {Curty}}, \ and\
  \bibinfo {author} {\bibfnamefont {K.}~\bibnamefont {Tamaki}},\ }\href
  {\doibase 10.1038/nphoton.2014.149} {\bibfield  {journal} {\bibinfo
  {journal} {Nat. Photonics}\ }\textbf {\bibinfo {volume} {8}},\ \bibinfo
  {pages} {595} (\bibinfo {year} {2014})}\BibitemShut {NoStop}%
\bibitem [{\citenamefont {Wehner}\ \emph {et~al.}(2018)\citenamefont {Wehner},
  \citenamefont {Elkouss},\ and\ \citenamefont {Hanson}}]{Wehner_2018}%
  \BibitemOpen
  \bibfield  {author} {\bibinfo {author} {\bibfnamefont {S.}~\bibnamefont
  {Wehner}}, \bibinfo {author} {\bibfnamefont {D.}~\bibnamefont {Elkouss}}, \
  and\ \bibinfo {author} {\bibfnamefont {R.}~\bibnamefont {Hanson}},\ }\href
  {\doibase 10.1126/science.aam9288} {\bibfield  {journal} {\bibinfo  {journal}
  {Science}\ }\textbf {\bibinfo {volume} {362}},\ \bibinfo {pages} {eaam9288}
  (\bibinfo {year} {2018})}\BibitemShut {NoStop}%
\bibitem [{\citenamefont {Pirker}\ and\ \citenamefont
  {Dür}(2019)}]{Pirker_2019}%
  \BibitemOpen
  \bibfield  {author} {\bibinfo {author} {\bibfnamefont {A.}~\bibnamefont
  {Pirker}}\ and\ \bibinfo {author} {\bibfnamefont {W.}~\bibnamefont {Dür}},\
  }\href {\doibase 10.1088/1367-2630/ab05f7} {\bibfield  {journal} {\bibinfo
  {journal} {New J. of Phys.}\ }\textbf {\bibinfo {volume} {21}},\ \bibinfo
  {pages} {033003} (\bibinfo {year} {2019})}\BibitemShut {NoStop}%
\bibitem [{\citenamefont {Azuma}\ \emph {et~al.}(2021)\citenamefont {Azuma},
  \citenamefont {Bäuml}, \citenamefont {Coopmans}, \citenamefont {Elkouss},\
  and\ \citenamefont {Li}}]{Azuma_2021}%
  \BibitemOpen
  \bibfield  {author} {\bibinfo {author} {\bibfnamefont {K.}~\bibnamefont
  {Azuma}}, \bibinfo {author} {\bibfnamefont {S.}~\bibnamefont {Bäuml}},
  \bibinfo {author} {\bibfnamefont {T.}~\bibnamefont {Coopmans}}, \bibinfo
  {author} {\bibfnamefont {D.}~\bibnamefont {Elkouss}}, \ and\ \bibinfo
  {author} {\bibfnamefont {B.}~\bibnamefont {Li}},\ }\href {\doibase
  10.1116/5.0024062} {\bibfield  {journal} {\bibinfo  {journal} {{AVS} Quantum
  Sci.}\ }\textbf {\bibinfo {volume} {3}},\ \bibinfo {pages} {014101} (\bibinfo
  {year} {2021})}\BibitemShut {NoStop}%
\bibitem [{\citenamefont {Cirac}\ \emph {et~al.}(1999)\citenamefont {Cirac},
  \citenamefont {Ekert}, \citenamefont {Huelga},\ and\ \citenamefont
  {Macchiavello}}]{CiracDistributed}%
  \BibitemOpen
  \bibfield  {author} {\bibinfo {author} {\bibfnamefont {J.~I.}\ \bibnamefont
  {Cirac}}, \bibinfo {author} {\bibfnamefont {A.~K.}\ \bibnamefont {Ekert}},
  \bibinfo {author} {\bibfnamefont {S.~F.}\ \bibnamefont {Huelga}}, \ and\
  \bibinfo {author} {\bibfnamefont {C.}~\bibnamefont {Macchiavello}},\ }\href
  {\doibase 10.1103/PhysRevA.59.4249} {\bibfield  {journal} {\bibinfo
  {journal} {Phys. Rev. A}\ }\textbf {\bibinfo {volume} {59}},\ \bibinfo
  {pages} {4249} (\bibinfo {year} {1999})}\BibitemShut {NoStop}%
\bibitem [{\citenamefont {Cacciapuoti}\ \emph {et~al.}(2020)\citenamefont
  {Cacciapuoti}, \citenamefont {Caleffi}, \citenamefont {Tafuri}, \citenamefont
  {Cataliotti}, \citenamefont {Gherardini},\ and\ \citenamefont
  {Bianchi}}]{Cacciapuoti2020}%
  \BibitemOpen
  \bibfield  {author} {\bibinfo {author} {\bibfnamefont {A.~S.}\ \bibnamefont
  {Cacciapuoti}}, \bibinfo {author} {\bibfnamefont {M.}~\bibnamefont
  {Caleffi}}, \bibinfo {author} {\bibfnamefont {F.}~\bibnamefont {Tafuri}},
  \bibinfo {author} {\bibfnamefont {F.~S.}\ \bibnamefont {Cataliotti}},
  \bibinfo {author} {\bibfnamefont {S.}~\bibnamefont {Gherardini}}, \ and\
  \bibinfo {author} {\bibfnamefont {G.}~\bibnamefont {Bianchi}},\ }\href
  {\doibase 10.1109/mnet.001.1900092} {\bibfield  {journal} {\bibinfo
  {journal} {{IEEE} Network}\ }\textbf {\bibinfo {volume} {34}},\ \bibinfo
  {pages} {137} (\bibinfo {year} {2020})}\BibitemShut {NoStop}%
\bibitem [{\citenamefont {Hayashi}\ and\ \citenamefont
  {Morimae}(2015)}]{Hayashi15}%
  \BibitemOpen
  \bibfield  {author} {\bibinfo {author} {\bibfnamefont {M.}~\bibnamefont
  {Hayashi}}\ and\ \bibinfo {author} {\bibfnamefont {T.}~\bibnamefont
  {Morimae}},\ }\href {\doibase 10.1103/PhysRevLett.115.220502} {\bibfield
  {journal} {\bibinfo  {journal} {Phys. Rev. Lett.}\ }\textbf {\bibinfo
  {volume} {115}},\ \bibinfo {pages} {220502} (\bibinfo {year}
  {2015})}\BibitemShut {NoStop}%
\bibitem [{\citenamefont {Sekatski}\ \emph {et~al.}(2020)\citenamefont
  {Sekatski}, \citenamefont {W\"olk},\ and\ \citenamefont
  {D\"ur}}]{Sekatski2020}%
  \BibitemOpen
  \bibfield  {author} {\bibinfo {author} {\bibfnamefont {P.}~\bibnamefont
  {Sekatski}}, \bibinfo {author} {\bibfnamefont {S.}~\bibnamefont {W\"olk}}, \
  and\ \bibinfo {author} {\bibfnamefont {W.}~\bibnamefont {D\"ur}},\ }\href
  {\doibase 10.1103/PhysRevResearch.2.023052} {\bibfield  {journal} {\bibinfo
  {journal} {Phys. Rev. Research}\ }\textbf {\bibinfo {volume} {2}},\ \bibinfo
  {pages} {023052} (\bibinfo {year} {2020})}\BibitemShut {NoStop}%
\bibitem [{\citenamefont {Kessler}\ \emph {et~al.}(2014)\citenamefont
  {Kessler}, \citenamefont {Lovchinsky}, \citenamefont {Sushkov},\ and\
  \citenamefont {Lukin}}]{Kessler2014}%
  \BibitemOpen
  \bibfield  {author} {\bibinfo {author} {\bibfnamefont {E.~M.}\ \bibnamefont
  {Kessler}}, \bibinfo {author} {\bibfnamefont {I.}~\bibnamefont {Lovchinsky}},
  \bibinfo {author} {\bibfnamefont {A.~O.}\ \bibnamefont {Sushkov}}, \ and\
  \bibinfo {author} {\bibfnamefont {M.~D.}\ \bibnamefont {Lukin}},\ }\href
  {\doibase 10.1103/PhysRevLett.112.150802} {\bibfield  {journal} {\bibinfo
  {journal} {Phys. Rev. Lett.}\ }\textbf {\bibinfo {volume} {112}},\ \bibinfo
  {pages} {150802} (\bibinfo {year} {2014})}\BibitemShut {NoStop}%
\bibitem [{\citenamefont {Eldredge}\ \emph {et~al.}(2018)\citenamefont
  {Eldredge}, \citenamefont {Foss-Feig}, \citenamefont {Gross}, \citenamefont
  {Rolston},\ and\ \citenamefont {Gorshkov}}]{Eldredge2018}%
  \BibitemOpen
  \bibfield  {author} {\bibinfo {author} {\bibfnamefont {Z.}~\bibnamefont
  {Eldredge}}, \bibinfo {author} {\bibfnamefont {M.}~\bibnamefont {Foss-Feig}},
  \bibinfo {author} {\bibfnamefont {J.~A.}\ \bibnamefont {Gross}}, \bibinfo
  {author} {\bibfnamefont {S.~L.}\ \bibnamefont {Rolston}}, \ and\ \bibinfo
  {author} {\bibfnamefont {A.~V.}\ \bibnamefont {Gorshkov}},\ }\href {\doibase
  10.1103/PhysRevA.97.042337} {\bibfield  {journal} {\bibinfo  {journal} {Phys.
  Rev. A}\ }\textbf {\bibinfo {volume} {97}},\ \bibinfo {pages} {042337}
  (\bibinfo {year} {2018})}\BibitemShut {NoStop}%
\bibitem [{\citenamefont {Eisert}\ \emph {et~al.}(2020)\citenamefont {Eisert},
  \citenamefont {Hangleiter}, \citenamefont {Walk}, \citenamefont {Roth},
  \citenamefont {Markham}, \citenamefont {Parekh}, \citenamefont {Chabaud},\
  and\ \citenamefont {Kashefi}}]{Eisert2020}%
  \BibitemOpen
  \bibfield  {author} {\bibinfo {author} {\bibfnamefont {J.}~\bibnamefont
  {Eisert}}, \bibinfo {author} {\bibfnamefont {D.}~\bibnamefont {Hangleiter}},
  \bibinfo {author} {\bibfnamefont {N.}~\bibnamefont {Walk}}, \bibinfo {author}
  {\bibfnamefont {I.}~\bibnamefont {Roth}}, \bibinfo {author} {\bibfnamefont
  {D.}~\bibnamefont {Markham}}, \bibinfo {author} {\bibfnamefont
  {R.}~\bibnamefont {Parekh}}, \bibinfo {author} {\bibfnamefont
  {U.}~\bibnamefont {Chabaud}}, \ and\ \bibinfo {author} {\bibfnamefont
  {E.}~\bibnamefont {Kashefi}},\ }\href {\doibase 10.1038/s42254-020-0186-4}
  {\bibfield  {journal} {\bibinfo  {journal} {Nat. Rev. Phys.}\ }\textbf
  {\bibinfo {volume} {2}},\ \bibinfo {pages} {382} (\bibinfo {year}
  {2020})}\BibitemShut {NoStop}%
\bibitem [{\citenamefont {B{\u{a}}descu}\ \emph {et~al.}(2019)\citenamefont
  {B{\u{a}}descu}, \citenamefont {O{\textquotesingle}Donnell},\ and\
  \citenamefont {Wright}}]{Bdescu2019}%
  \BibitemOpen
  \bibfield  {author} {\bibinfo {author} {\bibfnamefont {C.}~\bibnamefont
  {B{\u{a}}descu}}, \bibinfo {author} {\bibfnamefont {R.}~\bibnamefont
  {O{\textquotesingle}Donnell}}, \ and\ \bibinfo {author} {\bibfnamefont
  {J.}~\bibnamefont {Wright}},\ }\href@noop {} {\bibfield  {journal} {\bibinfo
  {journal} {\textit{Proceedings of the 51st Annual {ACM} {SIGACT} Symposium on
  Theory of Computing}}\ } (\bibinfo {year} {ACM, New York, 2019})}\BibitemShut
  {NoStop}%
\bibitem [{\citenamefont {Thinh}\ \emph {et~al.}(2020)\citenamefont {Thinh},
  \citenamefont {Dall{\textquotesingle}Arno},\ and\ \citenamefont
  {Scarani}}]{Thinh_2020}%
  \BibitemOpen
  \bibfield  {author} {\bibinfo {author} {\bibfnamefont {L.~P.}\ \bibnamefont
  {Thinh}}, \bibinfo {author} {\bibfnamefont {M.}~\bibnamefont
  {Dall{\textquotesingle}Arno}}, \ and\ \bibinfo {author} {\bibfnamefont
  {V.}~\bibnamefont {Scarani}},\ }\href {\doibase 10.22331/q-2020-09-11-320}
  {\bibfield  {journal} {\bibinfo  {journal} {Quantum}\ }\textbf {\bibinfo
  {volume} {4}},\ \bibinfo {pages} {320} (\bibinfo {year} {2020})}\BibitemShut
  {NoStop}%
\bibitem [{\citenamefont {Yu}\ \emph {et~al.}(2022)\citenamefont {Yu},
  \citenamefont {Shang},\ and\ \citenamefont {Gühne}}]{Yu_review2022}%
  \BibitemOpen
  \bibfield  {author} {\bibinfo {author} {\bibfnamefont {X.-D.}\ \bibnamefont
  {Yu}}, \bibinfo {author} {\bibfnamefont {J.}~\bibnamefont {Shang}}, \ and\
  \bibinfo {author} {\bibfnamefont {O.}~\bibnamefont {Gühne}},\ }\href
  {\doibase https://doi.org/10.1002/qute.202100126} {\bibfield  {journal}
  {\bibinfo  {journal} {Adv. Quantum Technol.}\ }\textbf {\bibinfo {volume}
  {5}},\ \bibinfo {pages} {2100126} (\bibinfo {year} {2022})}\BibitemShut
  {NoStop}%
\bibitem [{\citenamefont {Kliesch}\ and\ \citenamefont
  {Roth}(2021)}]{Kliesch2021}%
  \BibitemOpen
  \bibfield  {author} {\bibinfo {author} {\bibfnamefont {M.}~\bibnamefont
  {Kliesch}}\ and\ \bibinfo {author} {\bibfnamefont {I.}~\bibnamefont {Roth}},\
  }\href {\doibase 10.1103/PRXQuantum.2.010201} {\bibfield  {journal} {\bibinfo
   {journal} {PRX Quantum}\ }\textbf {\bibinfo {volume} {2}},\ \bibinfo {pages}
  {010201} (\bibinfo {year} {2021})}\BibitemShut {NoStop}%
\bibitem [{\citenamefont {Cramer}\ \emph {et~al.}(2010)\citenamefont {Cramer},
  \citenamefont {Plenio}, \citenamefont {Flammia}, \citenamefont {Somma},
  \citenamefont {Gross}, \citenamefont {Bartlett}, \citenamefont
  {Landon-Cardinal}, \citenamefont {Poulin},\ and\ \citenamefont
  {Liu}}]{Cramer2010}%
  \BibitemOpen
  \bibfield  {author} {\bibinfo {author} {\bibfnamefont {M.}~\bibnamefont
  {Cramer}}, \bibinfo {author} {\bibfnamefont {M.~B.}\ \bibnamefont {Plenio}},
  \bibinfo {author} {\bibfnamefont {S.~T.}\ \bibnamefont {Flammia}}, \bibinfo
  {author} {\bibfnamefont {R.}~\bibnamefont {Somma}}, \bibinfo {author}
  {\bibfnamefont {D.}~\bibnamefont {Gross}}, \bibinfo {author} {\bibfnamefont
  {S.~D.}\ \bibnamefont {Bartlett}}, \bibinfo {author} {\bibfnamefont
  {O.}~\bibnamefont {Landon-Cardinal}}, \bibinfo {author} {\bibfnamefont
  {D.}~\bibnamefont {Poulin}}, \ and\ \bibinfo {author} {\bibfnamefont {Y.-K.}\
  \bibnamefont {Liu}},\ }\href {https://doi.org/10.1038/ncomms1147} {\bibfield
  {journal} {\bibinfo  {journal} {Nat. Commun.}\ }\textbf {\bibinfo {volume}
  {1}} (\bibinfo {year} {2010})}\BibitemShut {NoStop}%
\bibitem [{\citenamefont {Haah}\ \emph {et~al.}(2017)\citenamefont {Haah},
  \citenamefont {Harrow}, \citenamefont {Ji}, \citenamefont {Wu},\ and\
  \citenamefont {Yu}}]{Haah_2017}%
  \BibitemOpen
  \bibfield  {author} {\bibinfo {author} {\bibfnamefont {J.}~\bibnamefont
  {Haah}}, \bibinfo {author} {\bibfnamefont {A.~W.}\ \bibnamefont {Harrow}},
  \bibinfo {author} {\bibfnamefont {Z.}~\bibnamefont {Ji}}, \bibinfo {author}
  {\bibfnamefont {X.}~\bibnamefont {Wu}}, \ and\ \bibinfo {author}
  {\bibfnamefont {N.}~\bibnamefont {Yu}},\ }\href {\doibase
  10.1109/TIT.2017.2719044} {\bibfield  {journal} {\bibinfo  {journal} {IEEE
  Trans. Inf. Theory}\ }\textbf {\bibinfo {volume} {63}},\ \bibinfo {pages}
  {5628} (\bibinfo {year} {2017})}\BibitemShut {NoStop}%
\bibitem [{\citenamefont {Flammia}\ and\ \citenamefont
  {Liu}(2011)}]{Flammia2011}%
  \BibitemOpen
  \bibfield  {author} {\bibinfo {author} {\bibfnamefont {S.~T.}\ \bibnamefont
  {Flammia}}\ and\ \bibinfo {author} {\bibfnamefont {Y.-K.}\ \bibnamefont
  {Liu}},\ }\href {\doibase 10.1103/PhysRevLett.106.230501} {\bibfield
  {journal} {\bibinfo  {journal} {Phys. Rev. Lett.}\ }\textbf {\bibinfo
  {volume} {106}},\ \bibinfo {pages} {230501} (\bibinfo {year}
  {2011})}\BibitemShut {NoStop}%
\bibitem [{\citenamefont {Wang}\ and\ \citenamefont
  {Hayashi}(2019)}]{Wang2019}%
  \BibitemOpen
  \bibfield  {author} {\bibinfo {author} {\bibfnamefont {K.}~\bibnamefont
  {Wang}}\ and\ \bibinfo {author} {\bibfnamefont {M.}~\bibnamefont {Hayashi}},\
  }\href {\doibase 10.1103/PhysRevA.100.032315} {\bibfield  {journal} {\bibinfo
   {journal} {Phys. Rev. A}\ }\textbf {\bibinfo {volume} {100}},\ \bibinfo
  {pages} {032315} (\bibinfo {year} {2019})}\BibitemShut {NoStop}%
\bibitem [{\citenamefont {Li}\ \emph {et~al.}(2019)\citenamefont {Li},
  \citenamefont {Han},\ and\ \citenamefont {Zhu}}]{Li19}%
  \BibitemOpen
  \bibfield  {author} {\bibinfo {author} {\bibfnamefont {Z.}~\bibnamefont
  {Li}}, \bibinfo {author} {\bibfnamefont {Y.-G.}\ \bibnamefont {Han}}, \ and\
  \bibinfo {author} {\bibfnamefont {H.}~\bibnamefont {Zhu}},\ }\href {\doibase
  10.1103/PhysRevA.100.032316} {\bibfield  {journal} {\bibinfo  {journal}
  {Phys. Rev. A}\ }\textbf {\bibinfo {volume} {100}},\ \bibinfo {pages}
  {032316} (\bibinfo {year} {2019})}\BibitemShut {NoStop}%
\bibitem [{\citenamefont {Zhu}\ and\ \citenamefont
  {Hayashi}(2019{\natexlab{a}})}]{Zhu1}%
  \BibitemOpen
  \bibfield  {author} {\bibinfo {author} {\bibfnamefont {H.}~\bibnamefont
  {Zhu}}\ and\ \bibinfo {author} {\bibfnamefont {M.}~\bibnamefont {Hayashi}},\
  }\href {\doibase 10.1103/PhysRevLett.123.260504} {\bibfield  {journal}
  {\bibinfo  {journal} {Phys. Rev. Lett.}\ }\textbf {\bibinfo {volume} {123}},\
  \bibinfo {pages} {260504} (\bibinfo {year} {2019}{\natexlab{a}})}\BibitemShut
  {NoStop}%
\bibitem [{\citenamefont {Zhu}\ and\ \citenamefont
  {Hayashi}(2019{\natexlab{b}})}]{Zhu2}%
  \BibitemOpen
  \bibfield  {author} {\bibinfo {author} {\bibfnamefont {H.}~\bibnamefont
  {Zhu}}\ and\ \bibinfo {author} {\bibfnamefont {M.}~\bibnamefont {Hayashi}},\
  }\href {\doibase 10.1103/PhysRevA.100.062335} {\bibfield  {journal} {\bibinfo
   {journal} {Phys. Rev. A}\ }\textbf {\bibinfo {volume} {100}},\ \bibinfo
  {pages} {062335} (\bibinfo {year} {2019}{\natexlab{b}})}\BibitemShut
  {NoStop}%
\bibitem [{\citenamefont {Zhu}\ and\ \citenamefont
  {Hayashi}(2019{\natexlab{c}})}]{Zhu19}%
  \BibitemOpen
  \bibfield  {author} {\bibinfo {author} {\bibfnamefont {H.}~\bibnamefont
  {Zhu}}\ and\ \bibinfo {author} {\bibfnamefont {M.}~\bibnamefont {Hayashi}},\
  }\href {\doibase 10.1103/PhysRevA.99.052346} {\bibfield  {journal} {\bibinfo
  {journal} {Phys. Rev. A}\ }\textbf {\bibinfo {volume} {99}},\ \bibinfo
  {pages} {052346} (\bibinfo {year} {2019}{\natexlab{c}})}\BibitemShut
  {NoStop}%
\bibitem [{\citenamefont {Yu}\ \emph {et~al.}(2019)\citenamefont {Yu},
  \citenamefont {Shang},\ and\ \citenamefont {Gühne}}]{Yu_2019}%
  \BibitemOpen
  \bibfield  {author} {\bibinfo {author} {\bibfnamefont {X.-D.}\ \bibnamefont
  {Yu}}, \bibinfo {author} {\bibfnamefont {J.}~\bibnamefont {Shang}}, \ and\
  \bibinfo {author} {\bibfnamefont {O.}~\bibnamefont {Gühne}},\ }\href
  {https://doi.org/10.1038/s41534-019-0226-z} {\bibfield  {journal} {\bibinfo
  {journal} {npj Quantum Inf.}\ }\textbf {\bibinfo {volume} {5}},\ \bibinfo
  {pages} {112} (\bibinfo {year} {2019})}\BibitemShut {NoStop}%
\bibitem [{\citenamefont {Hayashi}(2009)}]{Hayashi_2009}%
  \BibitemOpen
  \bibfield  {author} {\bibinfo {author} {\bibfnamefont {M.}~\bibnamefont
  {Hayashi}},\ }\href {\doibase 10.1088/1367-2630/11/4/043028} {\bibfield
  {journal} {\bibinfo  {journal} {New J. Phys.}\ }\textbf {\bibinfo {volume}
  {11}},\ \bibinfo {pages} {043028} (\bibinfo {year} {2009})}\BibitemShut
  {NoStop}%
\bibitem [{\citenamefont {Riera-S\`abat}\ \emph
  {et~al.}(2021{\natexlab{a}})\citenamefont {Riera-S\`abat}, \citenamefont
  {Sekatski}, \citenamefont {Pirker},\ and\ \citenamefont {D\"ur}}]{Riera1}%
  \BibitemOpen
  \bibfield  {author} {\bibinfo {author} {\bibfnamefont {F.}~\bibnamefont
  {Riera-S\`abat}}, \bibinfo {author} {\bibfnamefont {P.}~\bibnamefont
  {Sekatski}}, \bibinfo {author} {\bibfnamefont {A.}~\bibnamefont {Pirker}}, \
  and\ \bibinfo {author} {\bibfnamefont {W.}~\bibnamefont {D\"ur}},\ }\href
  {\doibase 10.1103/PhysRevLett.127.040502} {\bibfield  {journal} {\bibinfo
  {journal} {Phys. Rev. Lett.}\ }\textbf {\bibinfo {volume} {127}},\ \bibinfo
  {pages} {040502} (\bibinfo {year} {2021}{\natexlab{a}})}\BibitemShut
  {NoStop}%
\bibitem [{\citenamefont {Miguel-Ramiro}\ \emph {et~al.}(2022)\citenamefont
  {Miguel-Ramiro}, \citenamefont {Riera-S\`abat},\ and\ \citenamefont
  {D\"ur}}]{miguel2022improving}%
  \BibitemOpen
  \bibfield  {author} {\bibinfo {author} {\bibfnamefont {J.}~\bibnamefont
  {Miguel-Ramiro}}, \bibinfo {author} {\bibfnamefont {F.}~\bibnamefont
  {Riera-S\`abat}}, \ and\ \bibinfo {author} {\bibfnamefont {W.}~\bibnamefont
  {D\"ur}},\ }\href {\doibase 10.1103/PhysRevLett.129.190504} {\bibfield
  {journal} {\bibinfo  {journal} {Phys. Rev. Lett.}\ }\textbf {\bibinfo
  {volume} {129}},\ \bibinfo {pages} {190504} (\bibinfo {year}
  {2022})}\BibitemShut {NoStop}%
\bibitem [{\citenamefont {Liu}\ \emph {et~al.}(2002)\citenamefont {Liu},
  \citenamefont {Long}, \citenamefont {Tong},\ and\ \citenamefont
  {Li}}]{Liu2002}%
  \BibitemOpen
  \bibfield  {author} {\bibinfo {author} {\bibfnamefont {X.~S.}\ \bibnamefont
  {Liu}}, \bibinfo {author} {\bibfnamefont {G.~L.}\ \bibnamefont {Long}},
  \bibinfo {author} {\bibfnamefont {D.~M.}\ \bibnamefont {Tong}}, \ and\
  \bibinfo {author} {\bibfnamefont {F.}~\bibnamefont {Li}},\ }\href {\doibase
  10.1103/PhysRevA.65.022304} {\bibfield  {journal} {\bibinfo  {journal} {Phys.
  Rev. A}\ }\textbf {\bibinfo {volume} {65}},\ \bibinfo {pages} {022304}
  (\bibinfo {year} {2002})}\BibitemShut {NoStop}%
\bibitem [{\citenamefont {Bennett}\ \emph {et~al.}(1993)\citenamefont
  {Bennett}, \citenamefont {Brassard}, \citenamefont {Cr\'epeau}, \citenamefont
  {Jozsa}, \citenamefont {Peres},\ and\ \citenamefont
  {Wootters}}]{Bennett1993}%
  \BibitemOpen
  \bibfield  {author} {\bibinfo {author} {\bibfnamefont {C.~H.}\ \bibnamefont
  {Bennett}}, \bibinfo {author} {\bibfnamefont {G.}~\bibnamefont {Brassard}},
  \bibinfo {author} {\bibfnamefont {C.}~\bibnamefont {Cr\'epeau}}, \bibinfo
  {author} {\bibfnamefont {R.}~\bibnamefont {Jozsa}}, \bibinfo {author}
  {\bibfnamefont {A.}~\bibnamefont {Peres}}, \ and\ \bibinfo {author}
  {\bibfnamefont {W.~K.}\ \bibnamefont {Wootters}},\ }\href {\doibase
  10.1103/PhysRevLett.70.1895} {\bibfield  {journal} {\bibinfo  {journal}
  {Phys. Rev. Lett.}\ }\textbf {\bibinfo {volume} {70}},\ \bibinfo {pages}
  {1895} (\bibinfo {year} {1993})}\BibitemShut {NoStop}%
\bibitem [{\citenamefont {Bouwmeester}\ \emph {et~al.}(1997)\citenamefont
  {Bouwmeester}, \citenamefont {Pan}, \citenamefont {Mattle}, \citenamefont
  {Eibl}, \citenamefont {Weinfurter},\ and\ \citenamefont
  {Zeilinger}}]{Bouwmeester1997}%
  \BibitemOpen
  \bibfield  {author} {\bibinfo {author} {\bibfnamefont {D.}~\bibnamefont
  {Bouwmeester}}, \bibinfo {author} {\bibfnamefont {J.-W.}\ \bibnamefont
  {Pan}}, \bibinfo {author} {\bibfnamefont {K.}~\bibnamefont {Mattle}},
  \bibinfo {author} {\bibfnamefont {M.}~\bibnamefont {Eibl}}, \bibinfo {author}
  {\bibfnamefont {H.}~\bibnamefont {Weinfurter}}, \ and\ \bibinfo {author}
  {\bibfnamefont {A.}~\bibnamefont {Zeilinger}},\ }\href {\doibase
  10.1038/37539} {\bibfield  {journal} {\bibinfo  {journal} {Nature (London)}\
  }\textbf {\bibinfo {volume} {390}},\ \bibinfo {pages} {575} (\bibinfo {year}
  {1997})}\BibitemShut {NoStop}%
\bibitem [{\citenamefont {Werner}(1989)}]{Werner89}%
  \BibitemOpen
  \bibfield  {author} {\bibinfo {author} {\bibfnamefont {R.~F.}\ \bibnamefont
  {Werner}},\ }\href {\doibase 10.1103/PhysRevA.40.4277} {\bibfield  {journal}
  {\bibinfo  {journal} {Phys. Rev. A}\ }\textbf {\bibinfo {volume} {40}},\
  \bibinfo {pages} {4277} (\bibinfo {year} {1989})}\BibitemShut {NoStop}%
\bibitem [{\citenamefont {Horodecki}\ and\ \citenamefont
  {Horodecki}(1999)}]{horodecki}%
  \BibitemOpen
  \bibfield  {author} {\bibinfo {author} {\bibfnamefont {M.}~\bibnamefont
  {Horodecki}}\ and\ \bibinfo {author} {\bibfnamefont {P.}~\bibnamefont
  {Horodecki}},\ }\href {\doibase 10.1103/PhysRevA.59.4206} {\bibfield
  {journal} {\bibinfo  {journal} {Phys. Rev. A}\ }\textbf {\bibinfo {volume}
  {59}},\ \bibinfo {pages} {4206} (\bibinfo {year} {1999})}\BibitemShut
  {NoStop}%
\bibitem [{\citenamefont {Bennett}\ \emph
  {et~al.}(1996{\natexlab{a}})\citenamefont {Bennett}, \citenamefont
  {Brassard}, \citenamefont {Popescu}, \citenamefont {Schumacher},
  \citenamefont {Smolin},\ and\ \citenamefont {Wootters}}]{Bennett1996}%
  \BibitemOpen
  \bibfield  {author} {\bibinfo {author} {\bibfnamefont {C.~H.}\ \bibnamefont
  {Bennett}}, \bibinfo {author} {\bibfnamefont {G.}~\bibnamefont {Brassard}},
  \bibinfo {author} {\bibfnamefont {S.}~\bibnamefont {Popescu}}, \bibinfo
  {author} {\bibfnamefont {B.}~\bibnamefont {Schumacher}}, \bibinfo {author}
  {\bibfnamefont {J.~A.}\ \bibnamefont {Smolin}}, \ and\ \bibinfo {author}
  {\bibfnamefont {W.~K.}\ \bibnamefont {Wootters}},\ }\href {\doibase
  10.1103/PhysRevLett.76.722} {\bibfield  {journal} {\bibinfo  {journal} {Phys.
  Rev. Lett.}\ }\textbf {\bibinfo {volume} {76}},\ \bibinfo {pages} {722}
  (\bibinfo {year} {1996}{\natexlab{a}})}\BibitemShut {NoStop}%
\bibitem [{\citenamefont {Nielsen}\ and\ \citenamefont
  {Chuang}(2010)}]{nielsen_chuang_2010}%
  \BibitemOpen
  \bibfield  {author} {\bibinfo {author} {\bibfnamefont {M.~A.}\ \bibnamefont
  {Nielsen}}\ and\ \bibinfo {author} {\bibfnamefont {I.~L.}\ \bibnamefont
  {Chuang}},\ }\href@noop {} {\bibfield  {journal} {\bibinfo  {journal}
  {\textit{Quantum Computation and Quantum Information}}\ } (\bibinfo {year}
  {Cambridge University Press, Cambridge, 2010})}\BibitemShut {NoStop}%
\bibitem [{\citenamefont {Riera-S\`abat}\ \emph
  {et~al.}(2021{\natexlab{b}})\citenamefont {Riera-S\`abat}, \citenamefont
  {Sekatski}, \citenamefont {Pirker},\ and\ \citenamefont {D\"ur}}]{Riera2}%
  \BibitemOpen
  \bibfield  {author} {\bibinfo {author} {\bibfnamefont {F.}~\bibnamefont
  {Riera-S\`abat}}, \bibinfo {author} {\bibfnamefont {P.}~\bibnamefont
  {Sekatski}}, \bibinfo {author} {\bibfnamefont {A.}~\bibnamefont {Pirker}}, \
  and\ \bibinfo {author} {\bibfnamefont {W.}~\bibnamefont {D\"ur}},\ }\href
  {\doibase 10.1103/PhysRevA.104.012419} {\bibfield  {journal} {\bibinfo
  {journal} {Phys. Rev. A}\ }\textbf {\bibinfo {volume} {104}},\ \bibinfo
  {pages} {012419} (\bibinfo {year} {2021}{\natexlab{b}})}\BibitemShut
  {NoStop}%
\bibitem [{\citenamefont {Daboul}\ \emph {et~al.}(2003)\citenamefont {Daboul},
  \citenamefont {Wang},\ and\ \citenamefont {Sanders}}]{Daboul2003}%
  \BibitemOpen
  \bibfield  {author} {\bibinfo {author} {\bibfnamefont {J.}~\bibnamefont
  {Daboul}}, \bibinfo {author} {\bibfnamefont {X.}~\bibnamefont {Wang}}, \ and\
  \bibinfo {author} {\bibfnamefont {B.~C.}\ \bibnamefont {Sanders}},\ }\href
  {\doibase 10.1088/0305-4470/36/10/312} {\bibfield  {journal} {\bibinfo
  {journal} {J. Phys. A: Math. Gen.}\ }\textbf {\bibinfo {volume} {36}},\
  \bibinfo {pages} {2525} (\bibinfo {year} {2003})}\BibitemShut {NoStop}%
\bibitem [{\citenamefont {Horodecki}\ \emph {et~al.}(2009)\citenamefont
  {Horodecki}, \citenamefont {Horodecki}, \citenamefont {Horodecki},\ and\
  \citenamefont {Horodecki}}]{Horodecki09}%
  \BibitemOpen
  \bibfield  {author} {\bibinfo {author} {\bibfnamefont {R.}~\bibnamefont
  {Horodecki}}, \bibinfo {author} {\bibfnamefont {P.}~\bibnamefont
  {Horodecki}}, \bibinfo {author} {\bibfnamefont {M.}~\bibnamefont
  {Horodecki}}, \ and\ \bibinfo {author} {\bibfnamefont {K.}~\bibnamefont
  {Horodecki}},\ }\href {\doibase 10.1103/RevModPhys.81.865} {\bibfield
  {journal} {\bibinfo  {journal} {Rev. Mod. Phys.}\ }\textbf {\bibinfo {volume}
  {81}},\ \bibinfo {pages} {865} (\bibinfo {year} {2009})}\BibitemShut
  {NoStop}%
\bibitem [{\citenamefont {Vedral}\ \emph {et~al.}(1997)\citenamefont {Vedral},
  \citenamefont {Plenio}, \citenamefont {Rippin},\ and\ \citenamefont
  {Knight}}]{Vedral1997}%
  \BibitemOpen
  \bibfield  {author} {\bibinfo {author} {\bibfnamefont {V.}~\bibnamefont
  {Vedral}}, \bibinfo {author} {\bibfnamefont {M.~B.}\ \bibnamefont {Plenio}},
  \bibinfo {author} {\bibfnamefont {M.~A.}\ \bibnamefont {Rippin}}, \ and\
  \bibinfo {author} {\bibfnamefont {P.~L.}\ \bibnamefont {Knight}},\ }\href
  {\doibase 10.1103/PhysRevLett.78.2275} {\bibfield  {journal} {\bibinfo
  {journal} {Phys. Rev. Lett.}\ }\textbf {\bibinfo {volume} {78}},\ \bibinfo
  {pages} {2275} (\bibinfo {year} {1997})}\BibitemShut {NoStop}%
\bibitem [{\citenamefont {Bennett}\ \emph
  {et~al.}(1996{\natexlab{b}})\citenamefont {Bennett}, \citenamefont
  {DiVincenzo}, \citenamefont {Smolin},\ and\ \citenamefont
  {Wootters}}]{Bennett_hashing}%
  \BibitemOpen
  \bibfield  {author} {\bibinfo {author} {\bibfnamefont {C.~H.}\ \bibnamefont
  {Bennett}}, \bibinfo {author} {\bibfnamefont {D.~P.}\ \bibnamefont
  {DiVincenzo}}, \bibinfo {author} {\bibfnamefont {J.~A.}\ \bibnamefont
  {Smolin}}, \ and\ \bibinfo {author} {\bibfnamefont {W.~K.}\ \bibnamefont
  {Wootters}},\ }\href {\doibase 10.1103/PhysRevA.54.3824} {\bibfield
  {journal} {\bibinfo  {journal} {Phys. Rev. A}\ }\textbf {\bibinfo {volume}
  {54}},\ \bibinfo {pages} {3824} (\bibinfo {year}
  {1996}{\natexlab{b}})}\BibitemShut {NoStop}%
\bibitem [{\citenamefont {Pirandola}\ \emph {et~al.}(2017)\citenamefont
  {Pirandola}, \citenamefont {Laurenza}, \citenamefont {Ottaviani},\ and\
  \citenamefont {Banchi}}]{Pirandola_2017}%
  \BibitemOpen
  \bibfield  {author} {\bibinfo {author} {\bibfnamefont {S.}~\bibnamefont
  {Pirandola}}, \bibinfo {author} {\bibfnamefont {R.}~\bibnamefont {Laurenza}},
  \bibinfo {author} {\bibfnamefont {C.}~\bibnamefont {Ottaviani}}, \ and\
  \bibinfo {author} {\bibfnamefont {L.}~\bibnamefont {Banchi}},\ }\href
  {\doibase 10.1038/ncomms15043} {\bibfield  {journal} {\bibinfo  {journal}
  {Nat. Commun.}\ }\textbf {\bibinfo {volume} {8}},\ \bibinfo {pages} {15043}
  (\bibinfo {year} {2017})}\BibitemShut {NoStop}%
\bibitem [{\citenamefont {Deutsch}\ \emph {et~al.}(1996)\citenamefont
  {Deutsch}, \citenamefont {Ekert}, \citenamefont {Jozsa}, \citenamefont
  {Macchiavello}, \citenamefont {Popescu},\ and\ \citenamefont
  {Sanpera}}]{Deutsch1996}%
  \BibitemOpen
  \bibfield  {author} {\bibinfo {author} {\bibfnamefont {D.}~\bibnamefont
  {Deutsch}}, \bibinfo {author} {\bibfnamefont {A.}~\bibnamefont {Ekert}},
  \bibinfo {author} {\bibfnamefont {R.}~\bibnamefont {Jozsa}}, \bibinfo
  {author} {\bibfnamefont {C.}~\bibnamefont {Macchiavello}}, \bibinfo {author}
  {\bibfnamefont {S.}~\bibnamefont {Popescu}}, \ and\ \bibinfo {author}
  {\bibfnamefont {A.}~\bibnamefont {Sanpera}},\ }\href {\doibase
  10.1103/PhysRevLett.77.2818} {\bibfield  {journal} {\bibinfo  {journal}
  {Phys. Rev. Lett.}\ }\textbf {\bibinfo {volume} {77}},\ \bibinfo {pages}
  {2818} (\bibinfo {year} {1996})}\BibitemShut {NoStop}%
\bibitem [{\citenamefont {Dür}\ and\ \citenamefont
  {Briegel}(2007)}]{Duur_2007}%
  \BibitemOpen
  \bibfield  {author} {\bibinfo {author} {\bibfnamefont {W.}~\bibnamefont
  {Dür}}\ and\ \bibinfo {author} {\bibfnamefont {H.~J.}\ \bibnamefont
  {Briegel}},\ }\href {\doibase 10.1088/0034-4885/70/8/r03} {\bibfield
  {journal} {\bibinfo  {journal} {Rep. Prog. Phys.}\ }\textbf {\bibinfo
  {volume} {70}},\ \bibinfo {pages} {1381} (\bibinfo {year}
  {2007})}\BibitemShut {NoStop}%
\bibitem [{\citenamefont {Hostens}\ \emph {et~al.}(2006)\citenamefont
  {Hostens}, \citenamefont {Dehaene},\ and\ \citenamefont
  {De~Moor}}]{Hostens2006}%
  \BibitemOpen
  \bibfield  {author} {\bibinfo {author} {\bibfnamefont {E.}~\bibnamefont
  {Hostens}}, \bibinfo {author} {\bibfnamefont {J.}~\bibnamefont {Dehaene}}, \
  and\ \bibinfo {author} {\bibfnamefont {B.}~\bibnamefont {De~Moor}},\ }\href
  {\doibase 10.1103/PhysRevA.73.062337} {\bibfield  {journal} {\bibinfo
  {journal} {Phys. Rev. A}\ }\textbf {\bibinfo {volume} {73}},\ \bibinfo
  {pages} {062337} (\bibinfo {year} {2006})}\BibitemShut {NoStop}%
\bibitem [{\citenamefont {Pallister}\ \emph {et~al.}(2018)\citenamefont
  {Pallister}, \citenamefont {Linden},\ and\ \citenamefont
  {Montanaro}}]{Pallister18}%
  \BibitemOpen
  \bibfield  {author} {\bibinfo {author} {\bibfnamefont {S.}~\bibnamefont
  {Pallister}}, \bibinfo {author} {\bibfnamefont {N.}~\bibnamefont {Linden}}, \
  and\ \bibinfo {author} {\bibfnamefont {A.}~\bibnamefont {Montanaro}},\ }\href
  {\doibase 10.1103/PhysRevLett.120.170502} {\bibfield  {journal} {\bibinfo
  {journal} {Phys. Rev. Lett.}\ }\textbf {\bibinfo {volume} {120}},\ \bibinfo
  {pages} {170502} (\bibinfo {year} {2018})}\BibitemShut {NoStop}%
\bibitem [{\citenamefont {Chernoff}(1952)}]{Chernoff_1952}%
  \BibitemOpen
  \bibfield  {author} {\bibinfo {author} {\bibfnamefont {H.}~\bibnamefont
  {Chernoff}},\ }\href {\doibase 10.1214/aoms/1177729330} {\bibfield  {journal}
  {\bibinfo  {journal} {Ann. Math. Statist.}\ }\textbf {\bibinfo {volume}
  {23}},\ \bibinfo {pages} {493} (\bibinfo {year} {1952})}\BibitemShut
  {NoStop}%
\end{thebibliography}%

\end{document}